\documentclass[onecolumn,preprintnumbers,amsmath,amssymb,notitlepage]{revtex4}
\usepackage{graphicx}
\usepackage{epsfig}
\usepackage{amsmath}
\usepackage{amsfonts}
\usepackage{amssymb}
\usepackage{textcomp}
\usepackage{feynmf}

\begin{document}

\title{Electroweak Theory with a Minimal Length}

\author{Martin Kober}
\email{kober@fias.uni-frankfurt.de}
\email{kober@th.physik.uni-frankfurt.de}

\affiliation{Frankfurt Institute for Advanced Studies (FIAS),
Institut f\"ur Theoretische Physik, Johann Wolfgang Goethe-Universit\"at,
Ruth-Moufang-Strasse 1, 60438 Frankfurt am Main, Germany}
\date{\today}

\begin{abstract}
According to the introduction of a minimal length to quantum field theory which is directly related to
a generalized uncertainty principle the implementation of the gauge principle becomes much more intricated.
It has been shown in another paper how gauge theories have to be extended in general, if there is assumed the
existence of a minimal length. In this paper this generalization of the description of gauge theories is
applied to the case of Yang-Mills theories with gauge group $SU(N)$ to consider especially the application
to the electroweak theory as it appears in the standard model. The modifications of the lepton-, Higgs- and
gauge field sector of the extended Lagrangian of the electroweak theory maintaining local gauge invariance
under $SU(2)_L\otimes U(1)_Y$ transformations are investigated. There appear additional interaction terms between the leptons
or the Higgs particle respectively with the photon and the W- and Z-bosons as well as additional self-interaction
terms of these gauge bosons themselves. It is remarkable that in the quark sector where the full gauge group
of the standard model, $SU(3)_c \otimes SU(2)_L\otimes U(1)_Y$, has to be considered there arise coupling terms
between the gluons und the W- and Z-bosons which means that the electroweak theory is not separated from
quantum chromodynamics anymore.
\end{abstract}

\maketitle

\section{Introduction}

Any approach to a consistent formulation of a quantum theory of gravity seems to imply the existence of a minimal
length scale in nature. Through the introduction of a generalized uncertainty principle the consequence of such a minimal
length to relativistic quantum field theories can be described \cite{Maggiore:1993kv},\cite{Kempf:1994su},\cite{Hinrichsen:1995mf}.
The assumption of a generalized uncertainty principle corresponds to an extension of the usual Heisenbergian commutation relation
between the position operator and the momentum operator in quantum mechanics. This means that the commutator between the position
operator and the momentum operator is not independent of the position operator and the momentum operator anymore. If the commutator
depends on a function of the momentum operator, this implies that there exists an additional minimal uncertainty of position itself
and not only between position and momentum and this corresponds to the introduction of a minimal length. 
Of course, the extension of the commutation relation between the position operator and the momentum operator causes
a modification of the expression describing the momentum operator in position space. Inserting the modified
momentum operator to quantum theoretical field equations leads to a corresponding modification of the dynamics
they describe and in this sense to an extended formulation of these quantum field theories. There have been explored
many aspects of quantum mechanics and quantum field theory \cite{Kempf:1994su},\cite{Hinrichsen:1995mf},\cite{Kempf:1996ss},\cite{Kempf:1996nk},\cite{Lubo:1999xg},\cite{Chang:2001kn},\cite{Hossenfelder:2003jz},\cite{Harbach:2003qz},\cite{Harbach:2005yu},\cite{Bleicher:2010dg},\cite{Hossenfelder:2004up},\cite{Hossenfelder:2006cw},\cite{Hossenfelder:2007re},\cite{Bang:2006va},\cite{Buisseret:2010cw},\cite{Nozari:2010qy},\cite{Shibusa:2007ju},\cite{Percacci:2010af},\cite{Kim:2008kc},\cite{Matsuo:2005fb},\cite{Moayedi:2010vp}
as well as many issues concerning gravity
\cite{Maggiore:1993rv},\cite{Maggiore:1993zu},\cite{Scardigli:1999jh},\cite{Capozziello:1999wx},\cite{Crowell:1999up},\cite{Chang:2001bm},\cite{Kim:2006rx},\cite{Park:2007az},\cite{Scardigli:2007bw},\cite{Kim:2007hf},\cite{Bina:2007wj},\cite{Battisti:2007zg},\cite{Zhu:2008cg},\cite{Battisti:2008rv},\cite{Vakili:2008zg},\cite{Myung:2009gv},\cite{Myung:2009ur},\cite{Ali:2009zq},\cite{Farmany:2009zz},\cite{Li:2009zz},\cite{Bina:2010ir},\cite{Kim:2010wc},\cite{Ali:2011ap},\cite{Das:2008kaa},\cite{Vakili:2008tt},\cite{Kim:2007bx}
under incorporation of a generalized uncertainty principle. Another very important question concerning the incorporation
of a minimal length to quantum field theory is the formulation of realistic interaction theories as they are delivered
by the standard model of particle physics. Since all realistic quantum field theories containing a description of
interactions are formulated as gauge theories and the gauge principle is therefore a constitutive ingredient of these
theories, it is a precondition for the treatment of the interaction theories of the standard model under incorporation of a
minimal length to consider the relation between the generalized uncertainty principle on the one hand and the gauge principle
on the other hand. In \cite{Hossenfelder:2003jz} the concern of gauge invariance in the context of quantum electrodynamics with
a minimal length induced by a generalized uncertainty principle has already been treated. A general description of the
incorporation of the generalized uncertainty principle to gauge theories has been given in \cite{Kober:2010sj} and there have
been considered the special cases of quantum electrodynamics and the $SO(3,1)$ gauge description of gravity as well. It is
inevitable to perform a series expansion, because the representation of the momentum operator in position space contains an
infinite number of derivative terms, which all have to be replaced by covariant derivatives to maintain invariance under local
gauge transformations. In \cite{Kober:2010sj} there has been used a series expansion in the modification parameter of the
generalized uncertainty principle in contrast to \cite{Hossenfelder:2003jz} where has been used a series expansion in the
product of the gauge coupling constant and the modification parameter. A series expansion in the modification parameter implies
for abelian as well as non abelian gauge theories additional interaction terms between the matter field and the gauge field as
well as additional self-interaction terms of the gauge field already in a calculation to the first order.
In the present paper the consideration of gauge theories under incorporation of a generalized uncertainty principle
\cite{Kober:2010sj} shall be applied to another realistic quantum field theory, namely the electroweak theory
of the standard model \cite{Weinberg:1967tq},\cite{Weinberg:1996}. A precondition for the formulation of the electroweak
theory under incorporation of a generalized uncertainty principle is the consideration of general Yang-Mills gauge theories
with gauge group $SU(N)$ \cite{Yang:1954},\cite{Ramond:1990}, since the electroweak theory is based on gauge group
$SU(2)_L \otimes U(1)_Y$. It is a very important property of the application of the gauge principle to quantum field theories
under incorporation of a generalized uncertainty principle that the combination of two gauge groups leads to an intersection
of the corresponding gauge potentials because of the appearing products of covariant derivatives. This has not only the
consequence that the photon couples to the W- and Z-bosons in such a generalized version of the electroweak theory, but
it also causes a combination of the electroweak gauge potential with the gauge potential of quantum chromodynamics
in the presence of the full gauge group of the standard model, $SU(3)_c \otimes SU(2)_L \otimes U(1)_Y$, as it has to be
considered in the quark sector of the standard model.

The paper is structured as follows: First there is given a short repitition of the incorporation of the generalized
uncertainty principle to free field equations and the corresponding generalization of the gauge principle in general.
After this there is investigated the case of general Yang-Mills theories with gauge group $SU(N)$. This is the precondition
to consider the special realistic case of the electroweak interaction with gauge group $SU(2)_L\otimes U(1)_Y$ as it is
described in the standard model. There is investigated the generalized Lagrangian for the leptons and their interaction with
the gauge fields describing the photon, the Z- and the W-bosons as well as the corresponding generalized Lagrangian describing
these fields themselves. The extended Lagrangians reflect a much more complex interaction structure. The same holds for the
Higgs sector. Since all additional terms contain further derivative or gauge potential factors, the masses generated by
spontaneous symmetry breaking remain unchanged. After this there is considered the quark sector where arises the
mentioned intersection between the electroweak and the strong interaction which means that the gauge bosons of the
electroweak theory are coupled to the gluons. A short consideration of possible phenomenological consequences of the
presented theory is given at the end.

\section{Generalized Uncertainty Principle, Modified Dirac Equation and generalized gauge principle}

At the beginning there will be given a short review of the implementation of a minimal length to quantum field theory
by postulating a generalized uncertainty principle between the position operator and the momentum operator and its
incorporation to gauge field theories in general. If the generalization of the uncertainty principle shall lead to a
minimal length but not spoil translation invariance, in the relativistic case where the time coordinate is included
it has to be of the following shape:

\begin{equation}
\left[\hat x^\mu,\hat p_\nu\right]=i\delta^\mu_\nu \left[1+\beta \hat p^\rho \hat p_\rho\right]+2i\beta \hat p^\mu \hat p_\nu,
\label{generalized_uncertainty_principle}
\end{equation}
where 
$\beta$ denotes the parameter which determines the intensity of the modification of the uncertainty principle.
Usually there is assumed that $\beta$ is directly related to the Planck scale.
The generalized uncertainty principle leads to the following generalized uncertainty relation between position
and momentum, if it is considered for the index of the position operator being equal to the index of the
momentum operator:

$\Delta x^\mu \Delta p_\mu \geq \frac{1}{2}\left(1+\beta \Delta p^\rho \Delta p_\rho
+\beta \langle p^\rho \rangle \langle p_\rho \rangle\right)+i\left(\beta \Delta p^\mu
\Delta p_\mu+\beta\langle p^\mu \rangle \langle p_\mu \rangle\right)
=\frac{1}{2}\left(1+3\beta \Delta p^\mu \Delta p_\mu
+3\beta \langle p^\rho \rangle \langle p_\rho \rangle\right)$. This generalized uncertainty relation can be written as

\begin{equation}
\Delta p^\mu=\frac{\Delta x^\mu}{3\beta}\pm \sqrt{\left(\frac{\Delta x^\mu}{3\beta}\right)^2-\frac{1}{3\beta}
-\langle p^\rho \rangle \langle p_\rho \rangle},
\end{equation}
what leads to the minimal position uncertainty in Minkowski space:
$l_s=\Delta x^\mu_{min}=\sqrt{3\beta}\sqrt{1+3\beta\langle p^\rho \rangle \langle p_\rho \rangle}$,
from which the smallest length can be derived:

\begin{equation}
l_s=\sqrt{3\beta},
\end{equation}
if there is made the assumption that $\hbar=c=1$ as it is done throughout this paper. To obtain the representation of the
momentum operator in position space corresponding to the modified uncertainty principle ($\ref{generalized_uncertainty_principle}$)
there has to be performed a series expansion. A consideration to the first order yields the following representations of the
position operator and the momentum operator in position space:

\begin{equation}
\hat x^\mu=x^\mu \quad,\quad \hat p_\mu=-i\left(1-\beta \partial^{\rho}\partial_\rho\right)\partial_\mu
+\mathcal{O}\left(\beta^2\right).
\label{operators_position-space}
\end{equation}
Inserting the expression for the momentum operator in ($\ref{operators_position-space}$) to the Lagrangian 
corresponding to the Dirac equation yields

\begin{equation}
\mathcal{L}_D^{\prime}=\bar \psi \left[i\left(1-\beta \partial^\rho \partial_\rho\right) \gamma^\mu \partial_\mu-m\right]\psi,
\label{modified_Dirac_equation}
\end{equation}
where the $\gamma^\mu$ denote the Dirac matrices and $\bar \psi=\psi^{\dagger}\gamma^0$. If there is postulated invariance
with respect to local gauge transformations of a certain gauge group, also the additional derivatives appearing in
($\ref{modified_Dirac_equation}$) have to be replaced by covariant derivatives to maintain local gauge invariance,
since they also act on the local unitary transformation operator $U(x)$:

\begin{equation}
\left(\bf{1}-\beta \partial^{\rho}\partial_\rho\right)\partial_\mu 
\rightarrow \left({\bf 1}-\beta D^{\rho}D_\rho\right)D_\mu,
\label{derivatives}
\end{equation}
implying a transition $\mathcal{L}_D^{\prime}\rightarrow \mathcal{L}_m^{\prime}$ to the following generalized matter
Lagrangian containing the coupling of the matter field to the gauge field which exhibits a more complicated structure now:   

\begin{equation}
\mathcal{L}_{m}^{\prime}=\bar \psi \left[i\left({\bf 1}-\beta D^\rho D_\rho\right) \gamma^\mu D_\mu-m\right]\psi.
\label{modified_gauge_Dirac_Lagrangian}
\end{equation}
Since the additional term introduced by the generalized uncertainty principle transforms according to

\begin{eqnarray}
i \beta \bar \psi D^\rho D_\rho \gamma^\mu D_\mu \psi \rightarrow 
i \beta \bar \psi U^{\dagger}(x) U(x) D^\rho U^{\dagger}(x) U(x) D_\rho U^{\dagger}(x) 
\gamma^\mu U(x) D_\mu U^{\dagger}(x) U(x) \psi
=i \beta \bar \psi D^\rho D_\rho \gamma^\mu D_\mu \psi,
\label{symmetry}
\end{eqnarray}
the Lagrangian ($\ref{modified_gauge_Dirac_Lagrangian}$) is invariant under local gauge transformations. 
It has been used in ($\ref{symmetry}$) that $U^{\dagger}(x)U(x)={\bf 1}$ and that $U(x)$ either commutes with 
$\gamma^\mu$, if $U(x)$ does not provide a transformation referring to space-time, or $\gamma^\mu$ transforms
according to $\gamma^\mu \rightarrow U(x) \gamma^\mu U^{\dagger}(x)$.
Therefore a generalized covariant derivative in accordance with ($\ref{derivatives}$) can be introduced:

\begin{equation}
\mathcal{D}_\mu \equiv \left({\bf 1}-\beta D^{\rho}D_\rho\right)D_\mu
=\left[{\bf 1}-\beta\left(\partial^{\rho}{\bf 1}+iA^{\rho}\right)\left(\partial_\rho{\bf 1}+iA_\rho \right)\right]
\left(\partial_\mu{\bf 1}+iA_\mu\right),
\label{definition_generalized_covariant_derivative}
\end{equation}
which behaviour under local gauge transformations equals because of ($\ref{symmetry}$) to the one
of the usual covariant derivative meaning that

\begin{equation}
\mathcal{D}_\mu \rightarrow U(x)\mathcal{D}_\mu U^{\dagger}(x).
\label{transformation_generalized_covariant_derivative}
\end{equation}
Accordingly, there has also to be defined a generalized field strength tensor for the gauge field, which is done by
replacing the usual covariant derivative by the generalized covariant derivative ($\ref{definition_generalized_covariant_derivative}$)
in the corresponding commutator:

\begin{equation}
\mathcal{F}_{\mu\nu}=-i[\mathcal{D}_\mu, \mathcal{D}_\nu].
\label{definition_generalized_tensor}
\end{equation}
This general idea of the incorporation of the gauge principle to quantum field theories with a minimal length
as well as its manifestation in the special cases of electrodynamics and gravity have been considered in
\cite{Kober:2010sj}.

\section{Yang-Mills Theory under incorporation of a generalized gauge principle}

The generalized appearance of gauge field theories arising from the generalization of the uncertainty principle and the
corresponding modification of free field equations as it has been described above will now be regarded with respect to the
special case of Yang-Mills gauge groups as they are used within the formulation of the standard model to consider specifically
the electroweak theory in the next section. The Lagrangians for the gauge field and the matter field
being invariant under gauge transformations induced by the unitary operator $U(x)=\exp\left[i\alpha^a(x) T^a\right]$
will be calculated in terms of the gauge potential $A_\mu^a T^a$ to the first order in the modification parameter $\beta$.
The $T^a$ describe the generators of the group fulfilling the Lie-Algebra $[T^a,T^b]=if^{abc}T^c$, where the $f^{abc}$ denote
the structure constants of the group. In this special case of a Yang-Mills theory there can be built the following generalized
Lagrangian of the gauge field by using ($\ref{definition_generalized_covariant_derivative}$) and ($\ref{definition_generalized_tensor}$)
$\mathcal{L}_{YM g}^{\prime}=\frac{1}{4}{\text tr}\left[\mathcal{F}_{\mu\nu} \mathcal{F}^{\mu\nu}\right]$
and thus the complete Lagrangian of the modified Yang-Mills gauge theory reads

\begin{equation}
\mathcal{L}_{YM}^{\prime}=\bar\psi \left(i\gamma^\mu
\mathcal{D}_\mu-m\right)\psi-\frac{1}{4}{\text tr}\left[\mathcal{F}_{\mu\nu}
\mathcal{F}^{\mu\nu}\right],
\label{generalized_Lagrangian}
\end{equation}
if the gauge potential in ($\ref{definition_generalized_covariant_derivative}$) is identified with the Yang-Mills
potential with respect to ($\ref{generalized_Lagrangian}$).
The generalized Lagrangian ($\ref{generalized_Lagrangian}$) for a Yang-Mills gauge theory under incorporation of
a generalized uncertainty principle shall now be explored explicitly. If the matter sector
($\ref{modified_gauge_Dirac_Lagrangian}$) of the generalized Lagrangian ($\ref{generalized_Lagrangian}$) is calculated
to the first order in the modification parameter $\beta$ for a general $SU(N)$ gauge group,
one obtains the following expression:

\begin{eqnarray}
\mathcal{L}_{YM m}^{\prime}&=&
\bar \psi\left(i\gamma^\mu \mathcal{D}_\mu-m \right)\psi
=\bar \psi \left[i\left(1-\beta D^\rho D_\rho\right) \gamma^\mu D_\mu-m\right]\psi\nonumber\\
&=&\bar \psi\left\{i\gamma^\mu\left[1-\beta \left(\partial^{\rho}{\bf 1}+iA^{\rho a}T^a\right)
\left(\partial_\rho{\bf 1}+iA_\rho^a T^a \right)\right]
\left(\partial_\mu {\bf 1}+iA_\mu^{a}T^a\right)-m\right\}\psi\nonumber\\
&=&\bar \psi\left\{i\gamma^\mu\left[\partial_\mu {\bf 1}+iA_\mu^{a}T^{a}
-\beta \left(\partial^{\rho}\partial_{\rho}\partial_\mu{\bf 1}
+i\partial^\rho A_\rho^a T^a\partial_\mu+2iA_{\rho}^{a}T^{a}\partial^\rho \partial_\mu
\right.\right.\right.\nonumber\\
&&\left.\left.\left.
-A^{\rho a}A_\rho^{b} T^{a} T^{b} \partial_\mu
+i\partial^\rho \partial_\rho A_\mu^{a}T^{a}
+2i\partial_\rho A_\mu^{a}T^a \partial^\rho
+iA_\mu^{a}T^{a}\partial^\rho \partial_\rho
-\partial^\rho A_\rho^{a}A_\mu^{b}T^{a} T^{b}
\right.\right.\right.\nonumber\\
&&\left.\left.\left.
-2A_\rho^{a}\partial^\rho A_\mu^b T^{a} T^{b}-2A_\rho^{a}A_\mu^{b}T^{a} T^{b}\partial^\rho
-iA^{a \rho}A_\rho^{b}A_\mu^{c}T^{a}T^{b}T^{c}\right)\right]-m\right\}\psi.
\label{generalized_matter_Lagrangian_Yang-Mills}
\end{eqnarray}
This Lagrangian of course describes a much more complicated interaction structure between the matter field and the Yang-Mills
gauge field than the usual Lagrangian giving rise to new vertices where more than one gauge boson interact with the matter
field. To determine the Lagrangian of the gauge field sector explicitly, the shape of the generalized field strength tensor
$\mathcal{F}_{\mu\nu}$ defined in ($\ref{definition_generalized_tensor}$) has to be calculated first.
In \cite{Kober:2010sj} there has been shown that the generalized field strength tensor $\mathcal{F}_{\mu\nu}$ expressed by
the usual field strength tensor $F_{\mu\nu}$ and the usual covariant derivative $D_\mu$ to the first order in $\beta$ reads
in the general case without reference to a special gauge group

\begin{equation}
\mathcal{F}_{\mu\nu}=-i\left[\mathcal{D}_\mu,\mathcal{D}_\nu\right]=
F_{\mu\nu}-2\beta D^\rho D_\rho F_{\mu\nu}-\beta\left(D^\rho F_{\mu\rho}D_\nu-D^\rho F_{\nu\rho}D_\mu\right)
-\beta \left(F_{\mu\rho}D^{\rho}D_\nu-F_{\nu\rho}D^{\rho}D_\mu\right)+\mathcal{O}\left(\beta^2\right).
\label{generalized_tensor}
\end{equation}
The corresponding Lagrangian of the Yang-Mills gauge field reads accordingly

\begin{eqnarray}
\mathcal{L}_{YM g}^{\prime}&=&\frac{1}{4}
\textrm{tr}\left[{\mathcal{F}}_{\mu\nu}{\mathcal{F}}^{\mu\nu}\right]
=\frac{1}{4}\textrm{tr}\left\{F_{\mu\nu}F^{\mu\nu}-\beta\left[2 F_{\mu\nu}\left(D^\rho D_\rho F^{\mu\nu}\right)+
F_{\mu\nu}\left(F^{\mu\rho}D_{\rho}D^\nu-F^{\nu\rho}D_{\rho}D^\mu\right)+F_{\mu\nu}\left(D_\rho F^{\mu\rho}D^\nu\right.\right.\right.\nonumber\\
&&\left.\left.\left.-D_\rho F^{\nu\rho}D^\mu\right)
+2\left(D^\rho D_\rho F_{\mu\nu}\right)F^{\mu\nu}+\left(F_{\mu\rho}D^{\rho}D_\nu-F_{\nu\rho}D^{\rho}D_\mu\right)F^{\mu\nu}
+\left(D^\rho F_{\mu\rho}D_\nu-D^\rho F_{\nu\rho}D_\mu\right)F^{\mu\nu}\right]\right\}+\mathcal{O}\left(\beta^2\right)\nonumber\\
&=&\frac{1}{4}\textrm{tr}\left\{F_{\mu\nu}F^{\mu\nu}-\beta\left[2 F_{\mu\nu}\left(D^\rho D_\rho F^{\mu\nu}\right)+
2F_{\mu\nu}\left(F^{\mu\rho}D_{\rho}D^\nu\right)+2F_{\mu\nu}\left(D_\rho F^{\mu\rho}D^\nu\right)\right.\right.\nonumber\\
&&\left.\left.+2\left(D^\rho D_\rho F_{\mu\nu}\right)F^{\mu\nu}+2\left(F_{\mu\rho}D^{\rho}D_\nu\right) F^{\mu\nu}
+2\left(D^\rho F_{\mu\rho}D_\nu\right) F^{\mu\nu}\right]\right\}+\mathcal{O}\left(\beta^2\right).
\label{generalized_Lagrangian_gauge-field}
\end{eqnarray}
In the last step it has been used the antisymmetry of the field strength tensor, $F_{\mu\nu}=-F_{\nu\mu}$, which holds since
the structure constants of a Lie algebra fulfil $f^{abc}=-f^{bac}$. The special manifestation of ($\ref{generalized_tensor}$)
within Yang-Mills theories is more involved than the special case of electrodynamics considered in \cite{Kober:2010sj},
since the components of the field strength tensor do not commute with each other anymore.
If ($\ref{generalized_Lagrangian_gauge-field}$) is expressed by the potential $A_\mu^{a}T^a$, it reads as following:

\begin{eqnarray}
\mathcal{L}_{YM g}^{\prime}&=&\mathcal{L}_{YM g}+\frac{\beta}{2}{\rm tr}\left[\left(\partial_\mu A_{\nu}^{a}
-\partial_\nu A_{\mu}^{a}-f^{abc}A_\mu^b A_\nu^c\right)T^a
\left(\partial^\rho \partial_\rho+i\partial^\rho A_\rho^d T^d
+2iA_\rho^d T^d \partial^\rho
-A^{\rho e}A_\rho^d T^e T^d \right)\right. \nonumber\\ &&\left.
\quad\times \left(\partial^\mu A^{\nu f}-\partial^\nu A^{\mu f}-f^{fgh}A^{\mu g} A^{\nu h}\right)T^f
\right. \nonumber\\ &&\left.
+\left(\partial_\mu A_{\nu}^{a}-\partial_\nu A_{\mu}^{a}-f^{abc}A_\mu^b A_\nu^c\right)T^a
\left(\partial^\mu A^{\rho d}-\partial^\rho A^{\mu d}-f^{def}A^{\mu e} A^{\rho f}\right)T^d
\left(i\partial_\rho A^{\nu g} T^g-A_\rho^h A^{\nu g} T^h T^g\right)
\right. \nonumber\\ &&\left.
+\left(\partial_\mu A_{\nu}^{a}-\partial_\nu A_{\mu}^{a}-f^{abc}A_\mu^b A_\nu^c\right)T^a
\left(\partial_\rho+iA_\rho^d T^d \right)\left(\partial^\mu A^{\rho e}-\partial^\rho A^{\mu e}
-f^{efg}A^{\mu f} A^{\rho g}\right)T^e \left(iA^{\nu h}T^h\right) \right.\nonumber\\ &&\left.
+\left(\partial^\rho \partial_\rho+i\partial^\rho A_\rho^d T^d
+2iA_\rho^d T^d \partial^\rho
-A^{\rho e}A_\rho^d T^e T^d \right)\left(\partial^\mu A^{\nu f}
-\partial^\nu A^{\mu f}-f^{fgh}A^{\mu g} A^{\nu h}\right)T^f \right. \nonumber\\ &&\left.
\quad\times \left(\partial_\mu A_{\nu}^{a}-\partial_\nu A_{\mu}^{a}-f^{abc}A_\mu^b A_\nu^c\right)T^a
\right. \nonumber\\ &&\left.
+\left(\partial^\mu A^{\rho d}-\partial^\rho A^{\mu d}-f^{def}A^{\mu e} A^{\rho f}\right)T^d
\left(i\partial_\rho A^{\nu g} T^g-A_\rho^h A^{\nu g} T^h T^g\right)\left(\partial_\mu
A_{\nu}^{a}-\partial_\nu A_{\mu}^{a}-f^{abc}A_\mu^b A_\nu^c\right)T^a
\right. \nonumber\\ &&\left.
+\left(\partial_\rho+iA_\rho^d T^d  \right)\left(\partial^\mu A^{\rho e}-\partial^\rho A^{\mu e}
-f^{efg}A^{\mu f} A^{\rho g}\right)T^e\left(iA^{\nu h}T^h\right)\left(\partial_\mu A_{\nu}^{a}
-\partial_\nu A_{\mu}^{a}-f^{abc}A_\mu^b A_\nu^c\right)T^a\right],
\label{generalized_Lagrangian_gauge-field_potential}
\end{eqnarray}
where $\mathcal{L}_{YMg}$ describes the usual Yang-Mills Lagrangian of the gauge field

\begin{equation}
\mathcal{L}_{YM g}=\frac{1}{2}\partial^\mu A^{\nu a}\left(\partial_\mu A_\nu^a-\partial_\nu A_\mu^a \right)
-\frac{1}{2}f^{abc}A^{\mu b}A^{\nu c}\left(\partial_\mu A_\nu^a-\partial_\nu A_\mu^a\right)
+\frac{1}{4} f^{abe}f^{cde}A_\mu^a A_\nu^b A^{\mu c} A^{\nu d}.
\label{usual_Yang-Mills_Lagrangian}
\end{equation}
There have to be used the relations ${\rm tr}[T^a T^b]=\frac{1}{2}\delta^{ab}$, ${\rm tr}[T^a T^b T^c]=\frac{i}{4}f^{abc}$
and ${\rm tr}[T^a T^b T^c T^d]=\frac{1}{8}\left(\delta^{ab}\delta^{cd}-f^{abe}f^{ecd}\right)$ to transform the explicit
expression of the extended Yang-Mills Lagrangian ($\ref{generalized_Lagrangian_gauge-field}$),
($\ref{generalized_Lagrangian_gauge-field_potential}$) to its final form which can be written as follows:

\begin{equation}
\mathcal{L}_{YMg}^{\prime}=\mathcal{L}_{YMg}+\mathcal{L}_{YMg\beta}
=\mathcal{L}_{YM g}+\mathcal{L}_{\beta A_2}+\mathcal{L}_{\beta A_3}
+\mathcal{L}_{\beta A_4}+\mathcal{L}_{\beta A_5}+\mathcal{L}_{\beta A_6},
\label{generalized_Yang-Mills_Lagrangian}
\end{equation}
where the additional Lagrangians $\mathcal{L}_{\beta A_2}$, $\mathcal{L}_{\beta A_3}$, $\mathcal{L}_{\beta A_4}$, $\mathcal{L}_{\beta A_5}$, $\mathcal{L}_{\beta A_6}$
are defined according to

\begin{eqnarray}
\mathcal{L}_{\beta A_2}&=&\beta\left(\partial_\mu A_{\nu}^{a}\partial^\rho \partial_\rho \partial^\mu A^{\nu a}
-\partial_\mu A_{\nu}^{a}\partial^\rho \partial_\rho \partial^\nu A^{\mu a}\right),
\label{Lagrangian_A2}
\end{eqnarray}
\begin{eqnarray}
\mathcal{L}_{\beta A_3}&=&\frac{\beta}{4}\left[\left(-2f^{agh}\partial_\mu A_{\nu}^{a}
\partial^\rho \partial_\rho A^{\mu g} A^{\nu h}
-4f^{agh}\partial_\mu A_{\nu}^{a}\partial^\rho A^{\mu g} \partial_\rho A^{\nu h}
-2f^{agh}\partial_\mu A_{\nu}^{a}A^{\mu g} \partial^\rho \partial_\rho A^{\nu h}
+2f^{agh}\partial_\nu A_{\mu}^{a}\partial^\rho \partial_\rho A^{\mu g} A^{\nu h}
\right.\right. \nonumber\\ &&\left.\left.
+4f^{agh}\partial_\nu A_{\mu}^{a} \partial^\rho A^{\mu g} \partial_\rho A^{\nu h}
+2f^{agh}\partial_\nu A_{\mu}^{a}A^{\mu g} \partial^\rho \partial_\rho A^{\nu h}
-2f^{abc}A_\mu^b A_\nu^c\partial^\rho \partial_\rho \partial^\mu A^{\nu a}
+2f^{abc}A_\mu^b A_\nu^c\partial^\rho \partial_\rho \partial^\nu A^{\mu a}
\right)\right.\nonumber\\&&\left.
+f^{adf}\left(-2\partial_\mu A_{\nu}^{a}\partial^\rho A_\rho^d\partial^\mu A^{\nu f}
+2\partial_\mu A_{\nu}^{a}\partial^\rho A_\rho^d \partial^\nu A^{\mu f}
-4\partial_\mu A_{\nu}^{a} A_\rho^d \partial^\rho \partial^\mu A^{\nu f}
+4\partial_\mu A_{\nu}^{a} A_\rho^d \partial^\rho \partial^\nu A^{\mu f}
\right.\right. \nonumber\\ &&\left.\left.
-2\partial_\mu A_{\nu}^{a}\partial^\mu A^{\rho d}\partial_\rho A^{\nu f}
+2\partial_\mu A_{\nu}^{a}\partial^\rho A^{\mu d}\partial_\rho A^{\nu f}
+2\partial_\nu A_{\mu}^{a}\partial^\mu A^{\rho d}\partial_\rho A^{\nu f}
-2\partial_\nu A_{\mu}^{a}\partial^\rho A^{\mu d}\partial_\rho A^{\nu f}
\right.\right. \nonumber\\ &&\left.\left.
-\partial_\mu A_{\nu}^{a}\partial^\mu \partial_\rho A^{\rho d}A^{\nu f}
+\partial_\mu A_{\nu}^{a}\partial^\rho \partial_\rho A^{\mu d}A^{\nu f}
+\partial_\nu A_{\mu}^{a}\partial^\mu \partial_\rho A^{\rho d}A^{\nu f}
-\partial_\nu A_{\mu}^{a}\partial^\rho \partial_\rho A^{\mu d} A^{\nu f}
\right)\right],
\label{Lagrangian_A3}
\end{eqnarray}
\begin{eqnarray}
\mathcal{L}_{\beta A_4}&=&\frac{\beta}{4}\left[\left(2f^{abc}f^{agh}A_\mu^b A_\nu^c
\partial^\rho \partial_\rho A^{\mu g} A^{\nu h}
+2f^{abc}f^{agh}A_\mu^b A_\nu^c A^{\mu g} \partial^\rho \partial_\rho A^{\nu h}
+4f^{abc}f^{agh}A_\mu^b A_\nu^c \partial^\rho A^{\mu g} \partial_\rho A^{\nu h}
\right)\right.\nonumber\\&&\left.
+f^{adf}\left(f^{fgh}\partial_\mu A_{\nu}^{a} \partial^\rho A_\rho^d A^{\mu g} A^{\nu h}
+2f^{fgh}\partial_\mu A_{\nu}^{a} A_\rho^d \partial^\rho A^{\mu g} A^{\nu h}
+2f^{fgh}\partial_\mu A_{\nu}^{a} A_\rho^d A^{\mu g} \partial^\rho A^{\nu h}
-f^{fgh}\partial_\nu A_{\mu}^{a} \partial^\rho A_\rho^d A^{\mu g} A^{\nu h}
\right.\right. \nonumber\\ &&\left.\left.
-2f^{fgh}\partial_\nu A_{\mu}^{a} A_\rho^d \partial^\rho A^{\mu g} A^{\nu h}
-2f^{fgh}\partial_\nu A_{\mu}^{a} A_\rho^d A^{\mu g} \partial^\rho A^{\nu h}
+f^{abc}A_\mu^b A_\nu^c \partial^\rho A_\rho^d\partial^\mu A^{\nu f}
-f^{abc}A_\mu^b A_\nu^c \partial^\rho A_\rho^d\partial^\nu A^{\mu f}
\right.\right. \nonumber\\ &&\left.\left.
+2f^{abc}A_\mu^b A_\nu^c A_\rho^d \partial^\rho \partial^\mu A^{\nu f}
-2f^{abc}A_\mu^b A_\nu^c A_\rho^d \partial^\rho \partial^\nu A^{\mu f}
+f^{deg}\partial_\mu A_{\nu}^{a}A^{\mu e} A^{\rho g}\partial_\rho A^{\nu f}
-f^{deg}\partial_\nu A_{\mu}^{a}A^{\mu e} A^{\rho g}\partial_\rho A^{\nu f}
\right.\right. \nonumber\\ &&\left.\left.
+f^{abc}A_\mu^b A_\nu^c \partial^\mu A^{\rho d}\partial_\rho A^{\nu f}
-f^{abc}A_\mu^b A_\nu^c \partial^\rho A^{\mu d}\partial_\rho A^{\nu f}
+f^{dhg}\partial_\mu A_{\nu}^{a} \partial_\rho A^{\mu h} A^{\rho g}A^{\nu f}
+f^{dhg}\partial_\mu A_{\nu}^{a} A^{\mu h}\partial_\rho A^{\rho g}A^{\nu f}
\right.\right. \nonumber\\ &&\left.\left.
+f^{dhg}\partial_\mu A_{\nu}^{a} A^{\mu h} A^{\rho g}\partial_\rho A^{\nu f}
-f^{dhg}\partial_\nu A_{\mu}^{a}\partial_\rho A^{\mu h} A^{\rho g}A^{\nu f}
-f^{dhg}\partial_\nu A_{\mu}^{a}A^{\mu h}\partial_\rho A^{\rho g}A^{\nu f}
-f^{dhg}\partial_\nu A_{\mu}^{a}A^{\mu h} A^{\rho g}\partial_\rho A^{\nu f}
\right.\right. \nonumber\\ &&\left.\left.
+f^{abc}A_\mu^b A_\nu^c \partial^\mu \partial_\rho A^{\rho d}A^{\nu f}
+f^{abc}A_\mu^b A_\nu^c \partial^\mu A^{\rho d} \partial_\rho A^{\nu f}
-f^{abc}A_\mu^b A_\nu^c \partial^\rho \partial_\rho A^{\mu d}A^{\nu f}
-f^{abc}A_\mu^b A_\nu^c \partial^\rho A^{\mu d} \partial_\rho A^{\nu f}
\right)\right.\nonumber\\&&\left.
+\frac{1}{4}\left(-\partial_\mu A_{\nu}^{a} A^{\rho a}A_\rho^e \partial^\mu A^{\nu e}
+\partial_\mu A_{\nu}^{a} A^{\rho a}A_\rho^e \partial^\nu A^{\mu e}
+\partial_\nu A_{\mu}^{a} A^{\rho a}A_\rho^e \partial^\mu A^{\nu e}
-\partial_\nu A_{\mu}^{a} A^{\rho a}A_\rho^e \partial^\nu A^{\mu e}
\right.\right. \nonumber\\ &&\left.\left.
-\partial_\mu A_{\nu}^{a}\partial^\mu A^{\rho a}A_\rho^e A^{\nu e}
+\partial_\mu A_{\nu}^{a}\partial^\rho A^{\mu a}A_\rho^e A^{\nu e}
+\partial_\nu A_{\mu}^{a}\partial^\mu A^{\rho a}A_\rho^e A^{\nu e}
-\partial_\nu A_{\mu}^{a}\partial^\rho A^{\mu a}A_\rho^e A^{\nu e}
\right.\right. \nonumber\\ &&\left.\left.
-\partial_\mu A_{\nu}^{a} A_\rho^a \partial^\mu A^{\rho e}A^{\nu e}
+\partial_\mu A_{\nu}^{a} A_\rho^a \partial^\rho A^{\mu e}A^{\nu e}
+\partial_\nu A_\mu^a A_\rho^a \partial^\mu A^{\rho e} A^{\nu e}
-\partial_\nu A_\mu^a A_\rho^a \partial^\rho A^{\mu e} A^{\nu e}
\right.\right. \nonumber\\ &&\left.\left.
-\partial_\mu A_{\nu}^{e} A^{\rho a}A_\rho^a \partial^\mu A^{\nu e}
+\partial_\mu A_{\nu}^{e} A^{\rho a}A_\rho^a \partial^\nu A^{\mu e}
+\partial_\nu A_{\mu}^{e} A^{\rho a}A_\rho^a \partial^\mu A^{\nu e}
-\partial_\nu A_{\mu}^{e} A^{\rho a}A_\rho^a \partial^\nu A^{\mu e}
\right.\right. \nonumber\\ &&\left.\left.
-\partial_\mu A_{\nu}^{e}\partial^\mu A^{\rho a}A_\rho^a A^{\nu e}
+\partial_\mu A_{\nu}^{e}\partial^\rho A^{\mu a}A_\rho^a A^{\nu e}
+\partial_\nu A_{\mu}^{e}\partial^\mu A^{\rho a}A_\rho^a A^{\nu e}
-\partial_\nu A_{\mu}^{e}\partial^\rho A^{\mu a}A_\rho^a A^{\nu e}
\right.\right. \nonumber\\ &&\left.\left.
-\partial_\mu A_{\nu}^{e} A_\rho^a \partial^\mu A^{\rho a}A^{\nu e}
+\partial_\mu A_{\nu}^{e} A_\rho^a \partial^\rho A^{\mu a}A^{\nu e}
+\partial_\nu A_{\mu}^{e}A_\rho^a \partial^\mu A^{\rho a}A^{\nu e}
-\partial_\nu A_{\mu}^{e}A_\rho^a \partial^\rho A^{\mu a}A^{\nu e}
\right)\right.\nonumber\\&&\left.
-\frac{1}{4} f^{adi} f^{ief}\left(-\partial_\mu A_{\nu}^{a} A^{\rho d}A_\rho^e \partial^\mu A^{\nu f}
+\partial_\mu A_{\nu}^{a} A^{\rho d}A_\rho^e \partial^\nu A^{\mu f}
+\partial_\nu A_{\mu}^{a} A^{\rho d}A_\rho^e \partial^\mu A^{\nu f}
-\partial_\nu A_{\mu}^{a} A^{\rho d}A_\rho^e \partial^\nu A^{\mu f}
\right.\right. \nonumber\\ &&\left.\left.
-\partial_\mu A_{\nu}^{a}\partial^\mu A^{\rho d}A_\rho^e A^{\nu f}
+\partial_\mu A_{\nu}^{a}\partial^\rho A^{\mu d}A_\rho^e A^{\nu f}
+\partial_\nu A_{\mu}^{a}\partial^\mu A^{\rho d}A_\rho^e A^{\nu f}
-\partial_\nu A_{\mu}^{a}\partial^\rho A^{\mu d}A_\rho^e A^{\nu f}
\right.\right. \nonumber\\ &&\left.\left.
-\partial_\mu A_{\nu}^{a} A_\rho^d \partial^\mu A^{\rho e}A^{\nu f}
+\partial_\mu A_{\nu}^{a} A_\rho^d \partial^\rho A^{\mu e}A^{\nu f}
+\partial_\nu A_\mu^a A_\rho^d \partial^\mu A^{\rho e} A^{\nu f}
-\partial_\nu A_\mu^a A_\rho^d \partial^\rho A^{\mu e} A^{\nu f}
\right.\right. \nonumber\\ &&\left.\left.
-\partial_\mu A_{\nu}^{f} A^{\rho a}A_\rho^d \partial^\mu A^{\nu e}
+\partial_\mu A_{\nu}^{f} A^{\rho a}A_\rho^d \partial^\nu A^{\mu e}
+\partial_\nu A_{\mu}^{f} A^{\rho a}A_\rho^d \partial^\mu A^{\nu e}
-\partial_\nu A_{\mu}^{f} A^{\rho a}A_\rho^d \partial^\nu A^{\mu e}
\right.\right. \nonumber\\ &&\left.\left.
-\partial_\mu A_{\nu}^{f}\partial^\mu A^{\rho a}A_\rho^d A^{\nu e}
+\partial_\mu A_{\nu}^{f}\partial^\rho A^{\mu a}A_\rho^d A^{\nu e}
+\partial_\nu A_{\mu}^{f}\partial^\mu A^{\rho a}A_\rho^d A^{\nu e}
-\partial_\nu A_{\mu}^{f}\partial^\rho A^{\mu a}A_\rho^d A^{\nu e}
\right.\right. \nonumber\\ &&\left.\left.
-\partial_\mu A_{\nu}^{f} A_\rho^a \partial^\mu A^{\rho d}A^{\nu e}
+\partial_\mu A_{\nu}^{f} A_\rho^a \partial^\rho A^{\mu d}A^{\nu e}
+\partial_\nu A_{\mu}^{f}A_\rho^a \partial^\mu A^{\rho d}A^{\nu e}
-\partial_\nu A_{\mu}^{f}A_\rho^a \partial^\rho A^{\mu d}A^{\nu e}\right)\right],
\label{Lagrangian_A4}
\end{eqnarray}
\begin{eqnarray}
\mathcal{L}_{\beta A_5}&=&\frac{\beta}{8}\left[f^{adf}\left(-2 f^{abc}f^{fgh} A_\mu^b A_\nu^c \partial^\rho
A_\rho^d A^{\mu g} A^{\nu h}
-4f^{abc}f^{fgh}A_\mu^b A_\nu^c A_\rho^d \partial^\rho A^{\mu g} A^{\nu h}
-4f^{abc}f^{fgh}A_\mu^b A_\nu^c A_\rho^d A^{\mu g} \partial^\rho A^{\nu h}
\right.\right. \nonumber\\ &&\left.\left.
-2f^{abc}f^{deg}A_\mu^b A_\nu^c A^{\mu e} A^{\rho g}\partial_\rho A^{\nu f}
-2f^{abc}f^{dhg}A_\mu^b A_\nu^c \partial_\rho A^{\mu h} A^{\rho g}A^{\nu f}
-2f^{abc}f^{dhg}A_\mu^b A_\nu^c A^{\mu h}\partial_\rho A^{\rho g}A^{\nu f}
\right.\right. \nonumber\\ &&\left.\left.
-2f^{abc}f^{dhg}A_\mu^b A_\nu^c A^{\mu h} A^{\rho g}\partial_\rho A^{\nu f}
\right)\right.\nonumber\\&&\left.
+\frac{1}{2}\left(f^{egh}\partial_\mu A_{\nu}^{a} A^{\rho a}A_\rho^e A^{\mu g} A^{\nu h}
-f^{egh}\partial_\nu A_{\mu}^{a} A^{\rho a}A_\rho^e A^{\mu g} A^{\nu h}
+f^{abc}A_\mu^b A_\nu^c A^{\rho a}A_\rho^e \partial^\mu A^{\nu e}
-f^{abc}A_\mu^b A_\nu^c A^{\rho a}A_\rho^e \partial^\nu A^{\mu e}
\right.\right. \nonumber\\ &&\left.\left.
+f^{agh}\partial_\mu A_{\nu}^{a}A^{\mu g} A^{\rho h}A_\rho^e A^{\nu e}
-f^{agh}\partial_\nu A_{\mu}^{a}A^{\mu g} A^{\rho h}A_\rho^e A^{\nu e}
+f^{abc}A_\mu^b A_\nu^c \partial^\mu A^{\rho a}A_\rho^e A^{\nu e}
-f^{abc}A_\mu^b A_\nu^c \partial^\rho A^{\mu a}A_\rho^e A^{\nu e}
\right.\right. \nonumber\\ &&\left.\left.
+f^{ehg}\partial_\mu A_{\nu}^{a} A_\rho^a A^{\mu h} A^{\rho g}A^{\nu e}
-f^{ehg}\partial_\nu A_{\mu}^{a} A_\rho^a A^{\mu h} A^{\rho g}A^{\nu e}
+f^{abc}A_\mu^b A_\nu^c A_\rho^a \partial^\mu A^{\rho e}A^{\nu e}
-f^{abc}A_\mu^b A_\nu^c A_\rho^a \partial^\rho A^{\mu e}A^{\nu e}
\right.\right. \nonumber\\ &&\left.\left.
+f^{egh}\partial_\mu A_{\nu}^{e} A^{\rho a}A_\rho^a A^{\mu g} A^{\nu h}
-f^{egh}\partial_\nu A_{\mu}^{e} A^{\rho a}A_\rho^a A^{\mu g} A^{\nu h}
+f^{ebc}A_\mu^b A_\nu^c A^{\rho a}A_\rho^a \partial^\mu A^{\nu e}
-f^{ebc}A_\mu^b A_\nu^c A^{\rho a}A_\rho^a \partial^\nu A^{\mu e}
\right.\right. \nonumber\\ &&\left.\left.
+f^{agh}\partial_\mu A_{\nu}^{e}A^{\mu g} A^{\rho h}A_\rho^a A^{\nu e}
-f^{agh}\partial_\nu A_{\mu}^{e}A^{\mu g} A^{\rho h}A_\rho^a A^{\nu e}
+f^{ebc}A_\mu^b A_\nu^c \partial^\mu A^{\rho a}A_\rho^d A^{\nu e}
-f^{ebc}A_\mu^b A_\nu^c \partial^\rho A^{\mu a}A_\rho^d A^{\nu e}
\right.\right. \nonumber\\ &&\left.\left.
+f^{ahg} \partial_\mu A_{\nu}^{e} A_\rho^a A^{\mu h} A^{\rho g}A^{\nu e}
-f^{ahg}\partial_\nu A_{\mu}^{e} A_\rho^a A^{\mu h} A^{\rho g}A^{\nu e}
+f^{ebc}A_\mu^b A_\nu^c A_\rho^a \partial^\mu A^{\rho a}A^{\nu e}
-f^{ebc}A_\mu^b A_\nu^c A_\rho^a \partial^\rho A^{\mu a}A^{\nu e}
\right)\right.\nonumber\\&&\left.
-\frac{1}{2}f^{adi}f^{ief}\left(f^{fgh}\partial_\mu A_{\nu}^{a} A^{\rho d}A_\rho^e A^{\mu g} A^{\nu h}
-f^{fgh}\partial_\nu A_{\mu}^{a} A^{\rho d}A_\rho^e A^{\mu g} A^{\nu h}
+f^{abc}A_\mu^b A_\nu^c A^{\rho d}A_\rho^e \partial^\mu A^{\nu f}
-f^{abc}A_\mu^b A_\nu^c A^{\rho d}A_\rho^e \partial^\nu A^{\mu f}
\right.\right. \nonumber\\ &&\left.\left.
+f^{dgh}\partial_\mu A_{\nu}^{a}A^{\mu g} A^{\rho h}A_\rho^e A^{\nu f}
-f^{dgh}\partial_\nu A_{\mu}^{a}A^{\mu g} A^{\rho h}A_\rho^e A^{\nu f}
+f^{abc}A_\mu^b A_\nu^c \partial^\mu A^{\rho d}A_\rho^e A^{\nu f}
-f^{abc}A_\mu^b A_\nu^c \partial^\rho A^{\mu d}A_\rho^e A^{\nu f}
\right.\right. \nonumber\\ &&\left.\left.
+f^{ehg} \partial_\mu A_{\nu}^{a} A_\rho^d A^{\mu h} A^{\rho g}A^{\nu f}
-f^{ehg}\partial_\nu A_{\mu}^{a} A_\rho^d A^{\mu h} A^{\rho g}A^{\nu f}
+f^{abc}A_\mu^b A_\nu^c A_\rho^d \partial^\mu A^{\rho e}A^{\nu f}
-f^{abc}A_\mu^b A_\nu^c A_\rho^d \partial^\rho A^{\mu e}A^{\nu f}
\right.\right. \nonumber\\ &&\left.\left.
+f^{egh}\partial_\mu A_{\nu}^{f} A^{\rho a}A_\rho^d A^{\mu g} A^{\nu h}
-f^{egh}\partial_\nu A_{\mu}^{f} A^{\rho a}A_\rho^d A^{\mu g} A^{\nu h}
+f^{fbc}A_\mu^b A_\nu^c A^{\rho a}A_\rho^d \partial^\mu A^{\nu e}
-f^{fbc}A_\mu^b A_\nu^c A^{\rho a}A_\rho^d \partial^\nu A^{\mu e}
\right.\right. \nonumber\\ &&\left.\left.
+f^{agh}\partial_\mu A_{\nu}^{f}A^{\mu g} A^{\rho h}A_\rho^d A^{\nu e}
-f^{agh}\partial_\nu A_{\mu}^{f}A^{\mu g} A^{\rho h}A_\rho^d A^{\nu e}
+f^{fbc}A_\mu^b A_\nu^c \partial^\mu A^{\rho d}A_\rho^d A^{\nu e}
-f^{fbc}A_\mu^b A_\nu^c \partial^\rho A^{\mu d}A_\rho^d A^{\nu e}
\right.\right. \nonumber\\ &&\left.\left.
+f^{dhg} \partial_\mu A_{\nu}^{f} A_\rho^a A^{\mu h} A^{\rho g}A^{\nu e}
-f^{dhg}\partial_\nu A_{\mu}^{f} A_\rho^a A^{\mu h} A^{\rho g}A^{\nu e}
+f^{fbc}A_\mu^b A_\nu^c A_\rho^a \partial^\mu A^{\rho d}A^{\nu e}
-f^{fbc}A_\mu^b A_\nu^c A_\rho^a \partial^\rho A^{\mu d}A^{\nu e}\right)\right],
\label{Lagrangian_A5}
\end{eqnarray}
\begin{eqnarray}
\mathcal{L}_{\beta A_6}&=&\frac{\beta}{16}\left[\left(
-f^{abc}f^{egh}A_\mu^b A_\nu^c A^{\rho a}A_\rho^e A^{\mu g} A^{\nu h}
-f^{abc}f^{agh}A_\mu^b A_\nu^c A^{\mu g} A^{\rho h}A_\rho^e A^{\nu e}
-f^{abc}f^{ehg}A_\mu^b A_\nu^c A_\rho^a A^{\mu h} A^{\rho g}A^{\nu e}
\right.\right. \nonumber\\ &&\left.\left.
-f^{ebc}f^{egh}A_\mu^b A_\nu^c A^{\rho a}A_\rho^a A^{\mu g} A^{\nu h}
-f^{ebc}f^{agh}A_\mu^b A_\nu^c A^{\mu g} A^{\rho h}A_\rho^a A^{\nu e}
-f^{ebc}f^{ahg}A_\mu^b A_\nu^c A_\rho^a A^{\mu h} A^{\rho g}A^{\nu e}
\right)\right.\nonumber\\&&\left.
-f^{adi}f^{ief}\left(-f^{abc}f^{fgh}A_\mu^b A_\nu^c A^{\rho d}A_\rho^e A^{\mu g} A^{\nu h}
-f^{abc}f^{dgh}A_\mu^b A_\nu^c A^{\mu g} A^{\rho h}A_\rho^e A^{\nu f}
-f^{abc}f^{ehg}A_\mu^b A_\nu^c A_\rho^d A^{\mu h} A^{\rho g}A^{\nu f}
\right.\right. \nonumber\\ &&\left.\left.
-f^{fbc}f^{egh}A_\mu^b A_\nu^c A^{\rho a}A_\rho^d A^{\mu g} A^{\nu h}
-f^{fbc}f^{agh}A_\mu^b A_\nu^c A^{\mu g} A^{\rho h}A_\rho^d A^{\nu e}
-f^{fbc}f^{dhg}A_\mu^b A_\nu^c A_\rho^a A^{\mu h} A^{\rho g}A^{\nu e}\right)\right].
\label{Lagrangian_A6}
\end{eqnarray}

\section{The Generalized Gauge Principle in the Electroweak Theory}

\subsection{Lepton Sector of the Generalized Electroweak Theory}

To formulate the electroweak theory of the standard model with a minimal length the general consideration of Yang-Mills
theories under incorporation of a generalized uncertainty principle has to be applied to the spcial case of 
$SU(2)_L \otimes U(1)_Y$, the gauge group of the electroweak theory. In the usual electroweak theory local
gauge invariance under this group can be maintained in the lepton sector by introducing the following covariant derivative:

\begin{equation}
D_\mu=\partial_\mu+iA_\mu^a \tau^a+iB_\mu y\quad,\quad a=1...3,
\label{usual_covariant_derivative_electroweak}
\end{equation}
where the $\tau^a$ denote the generators of the $SU(2)_L$ and $y$ denotes the hypercharge. Since the neutrinos only
appear as left-handed neutrinos, the generators of the gauge group with respect to the neutrino-electron-dublett
look as follows:

\begin{equation}
\tau^a=g\left(\frac{1+\gamma_5}{4}\right)\sigma^a \quad,\quad a=1...3 \quad,\quad
y=g^{\prime}\left[\left(\frac{1+\gamma_5}{4}\right){\bf 1}
+\left(\frac{1-\gamma_5}{2}\right)\right],
\label{generators_electroweak_fermion}
\end{equation}
where the $\sigma^a$ denote the Pauli matrices with respect to the weak isospin space, {\bf 1} denotes
the identity matrix within the weak isospin space, $\gamma_5$ is defined as $\gamma^5=i\gamma^0 \gamma^1 \gamma^2 \gamma^3$
and thus $\left(\frac{1 \pm \gamma_5}{2}\right)$ describes the projection to the left-handed or right-handed part of
a Dirac spinor respectively. The hypercharge $y$ is related to the charge $q$ according to the Gell-Mann-Nishijima relation
which reads by using the representation ($\ref{generators_electroweak_fermion}$) as follows:

\begin{equation}
q=\frac{1}{g}\tau^3-\frac{1}{g^{\prime}}y=\left(\begin{matrix}0&0\\0&-1\end{matrix}\right).
\end{equation}
More elobarately the covariant derivative of the electroweak theory ($\ref{usual_covariant_derivative_electroweak}$)
can be expressed as

\begin{equation}
D_\mu=\partial^\mu+\frac{i}{\sqrt{2}}W^{\mu} \left(\tau^1-i\tau^2 \right)
+\frac{i}{\sqrt{2}}W^{\mu \dagger}\left(\tau^1+i\tau^2\right)+iZ^\mu\left(\tau^3 \cos \theta_W+y \sin \theta_W \right)
+iA^\mu \left(-\tau^3 \sin \theta_W+y\cos \theta_W \right),
\label{covariant_derivative_electroweak_explicite}
\end{equation}
if the gauge fields $W^\mu$, $W^{\mu \dagger}$, $Z^\mu$ and $A^\mu$ are defined as follows:

\begin{eqnarray}
W^\mu=\frac{1}{\sqrt{2}}\left(A_1^\mu+iA_2^\mu\right),\quad W^{\mu \dagger}=\frac{1}{\sqrt{2}}\left(A_1^\mu-iA_2^\mu\right),\quad
Z^\mu=\cos \theta_W A_3^\mu+\sin \theta_W B^\mu,\quad A^\mu=-\sin \theta_W A_3^\mu+\cos \theta_W B^\mu.\nonumber\\
\label{definition_WW*ZA}
\end{eqnarray}
$\theta_W$ denotes the Weinberg angle which describes the relation between the coupling constants $g$ and $g^{\prime}$
referring to the gauge groups $SU(2)_L$ and $U(1)_Y$ respectively and is accordingly defined by

\begin{equation}
\cos \theta_W=\frac{g}{\sqrt{g^2+g^{\prime 2}}}\quad,\quad \sin \theta_W=\frac{g^{\prime}}{\sqrt{g^2+g^{\prime 2}}}.
\end{equation} 
The Lagrangian of the generalized electroweak theory consists of four parts, if the quark sector is omitted

\begin{equation}
\mathcal{L}_{EW}^{\prime}=\mathcal{L}_{lepton}^{\prime}+\mathcal{L}_{gauge-field}^{\prime}
+\mathcal{L}_{Higgs}^{\prime}+\mathcal{L}_{Yukawa}^{\prime}.
\end{equation}
Since Yukawa coupling terms contain no derivative, they remain unchanged which means $\mathcal{L}_{Yukawa}^{\prime}=\mathcal{L}_{Yukawa}$.
The generalized covariant derivative containing the minimal length which general form
is given in ($\ref{definition_generalized_covariant_derivative}$) is for the special case of the electroweak
theory obtained by inserting the covariant derivative of the usual electroweak theory
($\ref{covariant_derivative_electroweak_explicite}$) to ($\ref{definition_generalized_covariant_derivative}$)
and reads accordingly

\begin{eqnarray}
\mathcal{D}_\mu&=&\left(1-\beta D^\rho D_\rho\right)D_\mu=\left[1-\beta \left(\partial^\rho+iA^{\rho a}\tau^a+iB^\rho y\right)
\left(\partial_\rho+iA_\rho^b \tau^b+i B_\rho y\right)\right]\left(\partial_\mu+iA_\mu^c \tau^c+iB_\mu y\right)\nonumber\\
&=&\left\{1-\beta\left[\partial^\rho+\frac{i}{\sqrt{2}}W^{\rho} \left(\tau^1-i\tau^2 \right)
+\frac{i}{\sqrt{2}}W^{\rho\dagger}\left(\tau^1+i\tau^2\right)
+iZ^\rho\left(\tau^3 \cos \theta_W+y\sin \theta_W \right)+iA^\rho \left(-\tau^3 \sin \theta_W+y\cos \theta_W \right)\right]
\right.\nonumber\\
&&\left.\times\left[\partial_\rho+\frac{i}{\sqrt{2}}W_{\rho} \left(\tau^1-i\tau^2
\right)+\frac{i}{\sqrt{2}}W_{\rho}^{\dagger}\left(\tau^1+i\tau^2\right)
+iZ_\rho\left(\tau^3 \cos \theta_W+y \sin \theta_W \right)+iA_\rho \left(-\tau^3 \sin \theta_W+y\cos \theta_W
\right)\right]\right\}
\nonumber\\
&&\times \left[\partial_\mu+\frac{i}{\sqrt{2}}W_{\mu} \left(\tau^1-i\tau^2
\right)+\frac{i}{\sqrt{2}}W_{\mu}^{\dagger}\left(\tau^1+i\tau^2\right)
+iZ_\mu\left(\tau^3 \cos \theta_W+y \sin \theta_W \right)+iA_\mu \left(-\tau^3 \sin \theta_W+y\cos \theta_W \right)\right].
\label{covariant_derivative_electroweak_extended}
\end{eqnarray}
The expression ($\ref{covariant_derivative_electroweak_extended}$) for the covariant derivative of the electroweak theory
with a generalized uncertainty principle to the first order in the modification parameter can be transformed to

\begin{eqnarray}
\mathcal{D}_\mu&=&\partial_\mu+\frac{i}{\sqrt{2}}W_{\mu}
\left(\tau^1-i\tau^2 \right)+\frac{i}{\sqrt{2}}W_{\mu}^{\dagger}\left(\tau^1+i\tau^2\right)
+iZ_\mu\left(\tau^3 \cos \theta_W
+y\sin \theta_W \right)+iA_\mu \left(-\tau^3 \sin \theta_W+y\cos \theta_W \right)\nonumber\\
&&-\beta \left[\partial^\rho \partial_\rho \partial_\mu+
\theta_\mu(W,W^{\dagger},Z,A)\left({\bf 1}_\tau-\tau^3\right)
+\lambda_\mu(W,W^{\dagger},Z,A)\left({\bf 1}_\tau+\tau^3\right)
+\chi_\mu(W,W^{\dagger},Z,A)\left(\tau^1-i\tau^2\right)\right.\nonumber\\&&\left.\quad\quad
+\phi_\mu(W,W^{\dagger},Z,A)\left(\tau^1+i\tau^2\right)
+\omega_\mu(A,Z)
\right],
\label{covariant_derivative_electroweak_extended_new}
\end{eqnarray}
where $\theta_\mu(W,W^{\dagger},Z,A)$, $\lambda_\mu(W,W^{\dagger},Z,A)$, $\chi_\mu(W,W^{\dagger},Z,A)$,
$\phi_\mu(W,W^{\dagger},Z,A)$ and $\omega_\mu(A,Z)$ are defined as

\begin{eqnarray}
\theta_\mu(W,W^{\dagger},Z,A)&=&
\frac{i\left(g^{\prime 2}-g^2\right)}{2g\sqrt{g^2+g^{\prime 2}}}\left(\partial^\rho \partial_\rho Z_\mu+2\partial^\rho Z_\mu \partial_\rho
+Z_\mu \partial^\rho \partial_\rho+\partial^\rho Z_\rho \partial_\mu+2 Z^\rho \partial_\rho \partial_\mu\right)
\nonumber\\&&
+\frac{ig^{\prime}}{\sqrt{g^2+g^{\prime 2}}}\left(\partial^\rho \partial_\rho A_\mu+2\partial^\rho A_\mu \partial_\rho
+A_\mu \partial^\rho \partial_\rho+\partial^\rho A_\rho \partial_\mu+2 A^\rho \partial_\rho \partial_\mu\right)
\nonumber\\&&
-\frac{g}{2}\left(\partial^\rho W_\rho W_\mu^{\dagger}+2W^{\rho} \partial_\rho W_\mu^{\dagger} \right)
-\frac{\left(g^{\prime 2}-g^{2}\right)^2}{4g\left(g^2+g^{\prime 2}\right)}
\left(\partial^\rho Z_\rho Z_\mu+2Z^\rho \partial_\rho Z_\mu\right)
-\frac{g g^{\prime 2}}{g^2+g^{\prime 2}}\left(\partial^\rho A_\rho A_\mu+2A^\rho \partial_\rho A_\mu\right)
\nonumber\\ &&
-\frac{g^{\prime}\left(g^{\prime 2}-g^{2}\right)}{2\left(g^2+g^{\prime 2}\right)}\left(\partial^\rho Z_\rho A_\mu+2Z^\rho \partial_\rho A_\mu\right)
-\frac{g^{\prime}\left(g^{\prime 2}-g^{2}\right)}{2\left(g^2+g^{\prime 2}\right)}\left(\partial^\rho A_\rho Z_\mu+2A^\rho \partial_\rho Z_\mu\right)
+\frac{1}{g}\Theta(W,W^{\dagger},Z,A)\partial_\mu
\nonumber\\&&
+\frac{i}{\sqrt{2}}\Xi(W,Z,A) W_\mu^{\dagger}+\frac{i\left(g^{\prime 2}-g^{2}\right)}{2g\sqrt{g^2+g^{\prime 2}}}\Theta(W,W^{\dagger},Z,A) Z_\mu
+\frac{ig^{\prime}}{\sqrt{g^2+g^{\prime 2}}}\Theta(W,W^{\dagger},Z,A) A_\mu,
\label{theta}
\end{eqnarray}
\begin{eqnarray}
\lambda_\mu(W,W^{\dagger},Z,A)&=&
\frac{i\sqrt{g^2+g^{\prime 2}}}{2g}\left(\partial^\rho \partial_\rho Z_\mu+2\partial^\rho Z_\mu \partial_\rho
+Z_\mu \partial^\rho \partial_\rho+\partial^\rho Z_\rho \partial_\mu+2 Z^\rho \partial_\rho \partial_\mu\right)
\nonumber\\&&
-\frac{g}{2}\left(\partial^\rho W_\rho^{\dagger} W_\mu+2W^{\dagger \rho}\partial_\rho W_\mu\right)
-\frac{g^2+g^{\prime 2}}{4g}\left(\partial^\rho Z_\rho Z_\mu+2Z^\rho \partial_\rho Z_\mu\right)
+\frac{1}{g}\Lambda(W,W^{\dagger},Z)\partial_\mu
\nonumber\\&&
+\frac{i}{\sqrt{2}}\Phi(W^{\dagger},Z,A) W_\mu+\frac{i\sqrt{g^2+g^{\prime 2}}}{2g}\Lambda(W,W^{\dagger},Z) Z_\mu,
\label{lambda}
\end{eqnarray}
\begin{eqnarray}
\chi_\mu(W,W^{\dagger},Z,A)&=&
\frac{i}{\sqrt{2}}\left(\partial^\rho \partial_\rho W_\mu+2\partial^\rho W_\mu \partial_\rho
+W_\mu \partial^\rho \partial_\rho+\partial^\rho W_\rho \partial_\mu+2 W^\rho \partial_\rho \partial_\mu\right)
-\frac{\sqrt{g^2+g^{\prime 2}}}{2\sqrt{2}}\left(\partial^\rho W_\rho Z_\mu+2W^\rho \partial_\rho Z_\mu\right)
\nonumber\\&&
-\frac{g^{\prime 2}-g^{2}}{2\sqrt{2\left(g^2+g^{\prime 2}\right)}}\left(\partial^\rho Z_\rho W_\mu+2 Z^\rho \partial_\rho W_\mu \right)
-\frac{g g^{\prime}}{\sqrt{2\left(g^2+g^{\prime 2}\right)}}\left(\partial^\rho A_\rho W_\mu
+2 A^\rho \partial_\rho W_\mu\right)
\nonumber\\ &&
+\frac{1}{g}\Xi(W,Z,A)\partial_\mu+\frac{i}{\sqrt{2}}\Theta(W,W^{\dagger},Z,A) W_\mu
+\frac{i\sqrt{g^2+g^{\prime 2}}}{2g}\Xi(W,Z,A) Z_\mu,
\label{chi}
\end{eqnarray}
\begin{eqnarray}
\phi_\mu(W,W^{\dagger},Z,A)&=&
\frac{i}{\sqrt{2}}\left(\partial^\rho \partial_\rho W_\mu^{\dagger}+2\partial^\rho W_\mu^{\dagger} \partial_\rho
+W_\mu^{\dagger} \partial^\rho \partial_\rho+\partial^\rho W_\rho^{\dagger} \partial_\mu+2 W^{\dagger \rho} \partial_\rho \partial_\mu\right)
-\frac{g^{\prime 2}-g^{2}}{2\sqrt{2\left(g^2+g^{\prime 2}\right)}}\left(\partial^\rho W_\rho^{\dagger} Z_\mu
+2 W^{\dagger \rho} \partial_\rho Z_\mu\right)
\nonumber\\&&
-\frac{\sqrt{g^2+g^{\prime 2}}}{2\sqrt{2}}\left(\partial^\rho Z_\rho W_\mu^{\dagger}+2Z^\rho \partial_\rho W_\mu^{\dagger}\right)
-\frac{g g^{\prime}}{\sqrt{2\left(g^2+g^{\prime 2}\right)}}\left(\partial^\rho W_\rho^{\dagger} A_\mu+2 W^{\dagger \rho}\partial_\rho A_\mu\right)
+\frac{1}{g}\Phi(W^{\dagger},Z,A)\partial_\mu
\nonumber\\ &&
+\frac{i}{\sqrt{2}}\Lambda(W,W^{\dagger},Z) W_\mu^{\dagger}
+\frac{i\left(g^{\prime 2}-g^{2}\right)}{2g\sqrt{g^2+g^{\prime 2}}}\Phi(W^{\dagger},Z,A) Z_\mu
+\frac{ig^{\prime}}{\sqrt{g^2+g^{\prime 2}}}\Phi(W^{\dagger},Z,A) A_\mu,
\label{phi}
\end{eqnarray}
\begin{eqnarray}
\omega_\mu(A,Z)&=&
\left(\frac{1-\gamma_5}{2}\right)\left[\frac{ig^{\prime 2}}{\sqrt{g^2+g^{\prime 2}}}
\left(\partial^\rho \partial_\rho Z_\mu+2\partial^\rho Z_\mu \partial_\rho
+Z_\mu \partial^\rho \partial_\rho+\partial^\rho Z_\rho \partial_\mu+2 Z^\rho \partial_\rho \partial_\mu\right)
\right.\nonumber\\&&\left.
+\frac{igg^{\prime}}{\sqrt{g^2+g^{\prime 2}}}
\left(\partial^\rho \partial_\rho A_\mu+2\partial^\rho A_\mu \partial_\rho
+A_\mu \partial^\rho \partial_\rho+\partial^\rho A_\rho \partial_\mu+2 A^\rho \partial_\rho \partial_\mu\right)
\right.\nonumber\\&&\left.
-\frac{g^{\prime 4}}{g^2+g^{\prime 2}}\left(\partial^\rho Z_\rho Z_\mu
+2Z^\rho \partial_\rho Z_\mu\right)
-\frac{g^{2}g^{\prime 2}}{g^2+g^{\prime 2}}
\left(\partial^\rho A_\rho A_\mu+2 A^\rho \partial_\rho A_\mu\right)
\right.\nonumber\\&&\left.
-\frac{g g^{\prime 3}}{g^2+g^{\prime 2}}
\left(\partial^\rho Z_\rho A_\mu+2 Z^\rho \partial_\rho A_\mu\right)
-\frac{g g^{\prime 3}}{g^2+g^{\prime 2}}
\left(\partial^\rho A_\rho Z_\mu+2 A^\rho \partial_\rho Z_\mu \right)
\right.\nonumber\\&&\left.
+\frac{ig^{\prime 2}}{\sqrt{g^2+g^{\prime 2}}}\Omega(A,Z) Z_\mu
+\frac{igg^{\prime}}{\sqrt{g^2+g^{\prime 2}}}\Omega(A,Z) A_\mu
+\Omega(A,Z)\partial_\mu \right],
\label{omega}
\end{eqnarray}
$\Theta(W,W^{\dagger},Z,A)$, $\Lambda(W,W^{\dagger},Z)$, $\Xi(W,Z,A)$,
$\Phi(W^{\dagger},Z,A)$ and $\Omega(A,Z)$ are defined as follows:

\begin{eqnarray}
\Theta(W,W^{\dagger},Z,A)&=&-\left[\frac{g^2}{2}\left(W^{\rho} W_\rho^{\dagger} \right)
+\frac{\left(g^{\prime 2}-g^{2}\right)^2}{4\left(g^2+g^{\prime 2}\right)}
\left(Z^\rho Z_\rho\right)+\frac{g^2 g^{\prime 2}}{g^2+g^{\prime 2}}\left(A^\rho A_\rho\right)
+\frac{gg^{\prime}\left(g^{\prime 2}-g^{2}\right)}{g^2+g^{\prime 2}}\left(Z^\rho A_\rho\right)
\right],
\nonumber\\
\Lambda(W,W^{\dagger},Z)&=&-\left[\frac{g^2}{2}\left(W^{\dagger \rho} W_\rho\right)
+\frac{g^2+g^{\prime 2}}{4}\left(Z^\rho Z_\rho\right)\right],
\nonumber\\
\Xi(W,Z,A)&=&-\left[\frac{g\sqrt{g^2+g^{\prime 2}}}{2\sqrt{2}}\left(W^\rho Z_\rho\right)
+\frac{g\left(g^{\prime 2}-g^{2}\right)}{2\sqrt{2\left(g^2+g^{\prime 2}\right)}}\left(Z^\rho W_{\rho}\right) 
+\frac{g^2 g^{\prime}}{\sqrt{2\left(g^2+g^{\prime 2}\right)}}\left(A^\rho W_\rho\right)\right],
\nonumber\\
\Phi(W^{\dagger},Z,A)&=&-\left[\frac{g\left(g^{\prime 2}-g^{2}\right)}{2\sqrt{2
\left(g^2+g^{\prime 2}\right)}}\left(W^{\dagger \rho} Z_\rho\right)
+\frac{g\sqrt{g^2+g^{\prime 2}}}{2\sqrt{2}}\left(Z^\rho W_\rho^{\dagger}\right)
+\frac{g^2 g^{\prime}}{\sqrt{2\left(g^2+g^{\prime 2}\right)}}
\left(W^{\dagger \rho} A_\rho\right)\right],
\nonumber\\
\Omega(A,Z)&=&-\left[\frac{g^{\prime 4}}{g^2+g^{\prime 2}}\left(Z^\rho Z_\rho\right)
+\frac{g^{2}g^{\prime 2}}{g^2+g^{\prime 2}}
\left(A^\rho A_\rho\right)
+\frac{2 g g^{\prime 3}}{g^2+g^{\prime 2}}\left(Z^\rho A_\rho\right)
\right],
\end{eqnarray}
and ${\bf 1}_\tau$ is defined as

\begin{eqnarray}
{\bf 1}_\tau=g\left(\frac{1+\gamma_5}{4}\right){\bf 1}.
\end{eqnarray}
To derive the expression ($\ref{covariant_derivative_electroweak_extended_new}$) there have to be used
the following identities:

\begin{equation}
 \left(\frac{1 \pm \gamma_5}{2}\right)^2=\left(\frac{1 \pm \gamma_5}{2}\right)\quad,\quad \left(\frac{1+\gamma_5}{2}\right)\left(\frac{1-\gamma_5}{2}\right)=0.
\end{equation}
The usual covariant derivative of the electroweak theory ($\ref{covariant_derivative_electroweak_explicite}$)
has to be replaced by the generalized one ($\ref{covariant_derivative_electroweak_extended_new}$)
within the lepton Langrangian to obtain the Lagrangian of the lepton sector in the presence of the
generalized uncertainty principle. This leads to the following expression for the lepton sector of the
generalized Lagrangian, if the electroweak theory is just considered with respect to the doublet
containing the electron-neutrino and the electron:

\begin{equation}
\mathcal{L}_{lepton}^{\prime}=i\left(\begin{matrix}\bar \nu_{e L} \ \bar e_L\end{matrix}\right)\gamma^\mu \mathcal{D}_\mu
\left(\begin{matrix}\nu_{e L} \\ e_L\end{matrix}\right)+i\bar e_R\gamma^\mu \mathcal{D}_\mu e_R,
\label{generalized_lepton_Lagrangian}
\end{equation}
where $e$ denotes the state of an electron, $\nu_e$ denotes the state of an electron-neutrino and it holds that

\begin{equation}
e_L=\left(\frac{1+\gamma_5}{2}\right)e\quad,\quad e_R=\left(\frac{1-\gamma_5}{2}\right)e
\quad,\quad \nu_{e L}=\gamma_5 \nu_{e L}.
\end{equation}
The Lagrangian for the other two lepton dubletts has of course completely the same structure as ($\ref{generalized_lepton_Lagrangian}$)
whereas the quark sector will be considered below, since this case is more complicated because of the incorporation of the
$SU(3)_c$ as gauge group of quantum chromodynamics which terms cannot be separated from the ones of the electroweak
Lagrangian anymore in the presence of a generalized uncertainty principle. If one expresses the generalized lepton
Lagrangian as the sum of the usual lepton Lagrangian and the modification to the first order in $\beta$

\begin{eqnarray}
\mathcal{L}_{lepton}^{\prime}&\equiv& \mathcal{L}_{lepton}+\mathcal{L}_{lepton\ \beta},
\label{generalized_lepton_Lagrangian_new}
\end{eqnarray}
where the usual lepton Lagrangian is defined as

\begin{eqnarray}
\mathcal{L}_{lepton}&=&i\bar \nu_e \gamma^\mu \mathcal{\partial}_\mu \nu_e+i\bar e\gamma^\mu \partial_\mu e
\nonumber\\&&
-\frac{g}{\sqrt{2}}\left[\bar e \gamma^\mu W_\mu \left(\frac{1+\gamma_5}{2}\right)\nu_e \right]
-\frac{g}{\sqrt{2}}\left[\bar \nu_e \gamma^\mu W_\mu^{\dagger}\left(\frac{1+\gamma_5}{2}\right)e\right]
-\frac{\sqrt{g^2+g^{{'}2}}}{2}\bar \nu_e \gamma^\mu Z_\mu \left(\frac{1+\gamma_5}{2}\right)\nu_e
\nonumber\\&&
+\frac{g^2-g^{\prime 2}}{2\sqrt{g^2+g^{\prime 2}}}\bar e \gamma^\mu Z_\mu \left(\frac{1+\gamma_5}{2}\right)e
-\frac{g^{\prime 2}}{\sqrt{g^2+g^{\prime 2}}}\bar e \gamma^\mu Z_\mu \left(\frac{1-\gamma_5}{2}\right)e
+\frac{gg^{\prime}}{\sqrt{g^2+g^{\prime 2}}}\left(\bar e \gamma^\mu A_\mu e\right),
\label{usual_lepton_Lagrangian}
\end{eqnarray}
then inserting of the expression of the generalized covariant derivative ($\ref{covariant_derivative_electroweak_extended_new}$)
into the lepton Lagrangian ($\ref{generalized_lepton_Lagrangian}$) leads to the following expression for the extension
of the lepton Lagrangian:

\begin{eqnarray}
\mathcal{L}_{lepton\ \beta}
&=&-i\beta\left[\bar \nu_e \gamma^\mu \partial^\rho \partial_\rho \partial_\mu \nu_e
+\bar e \gamma^\mu \partial^\rho \partial_\rho \partial_\mu e
+g\bar e\gamma^\mu
\theta_\mu(W^{\dagger},W,A,Z)\left(\frac{1+\gamma_5}{2}\right) e
+g\bar \nu_e \gamma^\mu
\lambda_\mu(W^{\dagger},W,A,Z)\left(\frac{1+\gamma_5}{2}\right) \nu_e \right.\nonumber\\&&\left.\quad\quad
+g\bar e\gamma^\mu
\chi_\mu(W^{\dagger},W,A,Z)\left(\frac{1+\gamma_5}{2}\right) \nu_e
+g\bar \nu_e \gamma^\mu
\phi_\mu(W^{\dagger},W,A,Z)\left(\frac{1+\gamma_5}{2}\right) e
+\bar e\gamma^\mu
\omega_\mu(A,Z) e\right],
\label{lepton_Lagrangian_beta}
\end{eqnarray}
where $\theta_\mu(W,W^{\dagger},Z,A)$, $\lambda_\mu(W,W^{\dagger},Z,A)$, $\phi_\mu(W,W^{\dagger},Z,A)$,
$\chi_\mu(W,W^{\dagger},Z,A)$ and $\omega_\mu(Z,A)$ are defined according to ($\ref{theta}$),($\ref{lambda}$),($\ref{chi}$),($\ref{phi}$)
and ($\ref{omega}$). Thus there has been constructed a lepton Lagrangian, ($\ref{generalized_lepton_Lagrangian_new}$)
with  ($\ref{usual_lepton_Lagrangian}$) and ($\ref{lepton_Lagrangian_beta}$), containing the generalized uncertainty
principle and being invariant under local gauge transformations of the form

\begin{eqnarray}
\left(\begin{matrix} \nu_{e L}\\ e_L \end{matrix}\right)&\rightarrow& \exp\left[i\alpha_a(x) \tau^a+i\alpha_y(x) y\right]
\left(\begin{matrix} \nu_{e L} \\ e_L \end{matrix}\right)
\quad,\quad  e_R \rightarrow \exp\left[i\alpha_y(x) y\right] e_R,
\nonumber\\ \mathcal{D}_\mu &\rightarrow& \exp\left[i\alpha_a(x) \tau^a+i\alpha_y(x) y\right]\mathcal{D}_\mu
\exp\left[-i\alpha_a(x) \tau^a-i\alpha_y(x) y\right]
\quad,\quad a=1...3.
\end{eqnarray}

\subsection{Higgs Sector of the Generalized Electroweak Theory}

To formulate the extended Lagrangian of the Higgs sector there is used the following
representation of the generators of the gauge group of the electroweak theory:

\begin{equation}
\tau_\varphi^a=\frac{g}{2}\sigma^a \quad,\quad y_\varphi=-\frac{g^{\prime}}{2}{\bf 1}.
\label{generators_phi}
\end{equation}
As consequence the charge operator has the following form:

\begin{equation}
q_\varphi=\frac{1}{g}\tau_\varphi^3-\frac{1}{g^{\prime}}y_\varphi=\left(\begin{matrix}1&0\\0&0\end{matrix}\right).
\label{charge_isospin_hypercharge}
\end{equation}
Of course, ${\bf 1}_{\tau \varphi}$ is accordingly defined as ${\bf 1}_{\tau \varphi}=\frac{g}{2}{\bf 1}$.
The generalized covariant derivative with respect to the generators ($\ref{generators_phi}$) reads

\begin{eqnarray}
\mathcal{D}_\mu&=&\partial_\mu+\frac{i}{\sqrt{2}}W_{\mu}
\left(\tau_{\varphi}^1-i\tau_{\varphi}^2 \right)+\frac{i}{\sqrt{2}}W_{\mu}^{\dagger}
\left(\tau_{\varphi}^1+i\tau_{\varphi}^2\right)+iZ_\mu\left(\tau_{\varphi}^3 \cos \theta_W
+y_{\varphi}\sin \theta_W \right)+iA_\mu \left(-\tau_{\varphi}^3 \sin \theta_W
+y_{\varphi}\cos \theta_W \right)\nonumber\\
&&-\beta \left[\partial^\rho \partial_\rho \partial_\mu
+\theta_{\varphi \mu}(W,W^{\dagger},Z,A)\left({\bf 1}_{\tau \varphi}-\tau_{\varphi}^3\right)
+\lambda_{\varphi \mu}(W,W^{\dagger},Z,A)\left({\bf 1}_{\tau \varphi}+\tau_{\varphi}^3\right)
+\chi_{\varphi \mu}(W,W^{\dagger},Z,A)\left(\tau_{\varphi}^1-i\tau_{\varphi}^2\right)
\right.\nonumber\\&&\left.\quad\quad
+\phi_{\varphi \mu}(W,W^{\dagger},Z,A)\left(\tau_{\varphi}^1+i\tau_{\varphi}^2\right)\right],
\label{covariant_derivative_electroweak_extended_Higgs}
\end{eqnarray}
where $\theta_{\varphi \mu}(W,W^{\dagger},Z,A)$, $\lambda_{\varphi \mu}(W,W^{\dagger},Z,A)$,
$\chi_{\varphi \mu}(W,W^{\dagger},Z,A)$ and $\phi_{\varphi \mu}(W,W^{\dagger},Z,A)$ are
defined as

\begin{eqnarray}
\theta_{\varphi \mu}(W,W^{\dagger},Z,A)&=&
-\frac{i\sqrt{g^2+g^{\prime 2}}}{2g}
\left(\partial^\rho \partial_\rho Z_\mu+2\partial^\rho Z_\mu \partial_\rho
+Z_\mu \partial^\rho \partial_\rho+\partial^\rho Z_\rho \partial_\mu
+2 Z^\rho \partial_\rho \partial_\mu\right)
\nonumber\\&&
-\frac{g}{2}\left(\partial^\rho W_\rho W_\mu^{\dagger}+2W^{\rho} \partial_\rho W_\mu^{\dagger}\right)
-\frac{g^2+g^{\prime 2}}{4g}
\left(\partial^\rho Z_\rho Z_\mu+2Z^\rho \partial_\rho Z_\mu\right)
+\frac{1}{g}\Theta_\varphi(W,W^{\dagger},Z)\partial_\mu
\nonumber\\&&
+\frac{i}{\sqrt{2}}\Xi_\varphi(W,Z,A) W_\mu^{\dagger}
-\frac{i\sqrt{g^{2}+g^{\prime 2}}}{2g}\Theta_\varphi(W,W^{\dagger},Z) Z_\mu,
\label{theta_Higgs}
\end{eqnarray}
\begin{eqnarray}
\lambda_{\varphi \mu}(W,W^{\dagger},Z,A)&=&
\frac{i\left(g^{2}-g^{\prime 2}\right)}{2g\sqrt{g^2+g^{\prime 2}}}
\left(\partial^\rho \partial_\rho Z_\mu+2\partial^\rho Z_\mu \partial_\rho
+Z_\mu \partial^\rho \partial_\rho+\partial^\rho Z_\rho \partial_\mu+2 Z^\rho \partial_\rho \partial_\mu\right)
\nonumber\\&&
-\frac{ig^{\prime}}{\sqrt{g^2+g^{\prime 2}}}\left(\partial^\rho \partial_\rho A_\mu
+2\partial^\rho A_\mu \partial_\rho+A_\mu \partial^\rho \partial_\rho
+\partial^\rho A_\rho \partial_\mu+2 A^\rho \partial_\rho \partial_\mu\right)
\nonumber\\&&
-\frac{g}{2}\left(\partial^\rho W_\rho^{\dagger} W_\mu+2W^{\dagger \rho}\partial_\rho W_\mu\right)
-\frac{\left(g^{2}-g^{\prime 2}\right)^2}{4g\left(g^2+g^{\prime 2}\right)}
\left(\partial^\rho Z_\rho Z_\mu+2Z^\rho \partial_\rho Z_\mu\right)
-\frac{g g^{\prime 2}}{g^2+g^{\prime 2}}\left(\partial^\rho A_\rho A_\mu+2A^\rho \partial_\rho A_\mu\right)
\nonumber\\ &&
+\frac{g^{\prime}\left(g^{2}-g^{\prime 2}\right)}{2\left(g^2+g^{\prime 2}\right)}
\left(\partial^\rho Z_\rho A_\mu+2Z^\rho \partial_\rho A_\mu\right)
+\frac{g^{\prime}\left(g^{2}-g^{\prime 2}\right)}{2\left(g^2+g^{\prime 2}\right)}
\left(\partial^\rho A_\rho Z_\mu+2A^\rho \partial_\rho Z_\mu\right)
+\frac{1}{g}\Lambda_\varphi(W,W^{\dagger},Z,A)\partial_\mu
\nonumber\\&&
+\frac{i}{\sqrt{2}}\Phi_\varphi(W^{\dagger},Z,A) W_\mu
+\frac{i\left(g^2-g^{\prime 2}\right)}{2g\sqrt{g^2+g^{\prime 2}}}\Lambda_\varphi(W,W^{\dagger},Z,A) Z_\mu
-\frac{ig^{\prime}}{\sqrt{g^2+g^{\prime 2}}}\Lambda_\varphi(W,W^{\dagger},Z,A) A_\mu,
\label{lambda_Higgs}
\end{eqnarray}
\begin{eqnarray}
\chi_{\varphi \mu}(W,W^{\dagger},Z,A)&=&
\frac{i}{\sqrt{2}}\left(\partial^\rho \partial_\rho W_\mu+2\partial^\rho W_\mu \partial_\rho
+W_\mu \partial^\rho \partial_\rho+\partial^\rho W_\rho \partial_\mu+2 W^\rho \partial_\rho \partial_\mu\right)
-\frac{g^{2}-g^{\prime 2}}{2\sqrt{2\left(g^2+g^{\prime 2}\right)}}
\left(\partial^\rho W_\rho Z_\mu+2W^\rho \partial_\rho Z_\mu\right)
\nonumber\\&&
+\frac{\sqrt{g^2+g^{\prime 2}}}{2\sqrt{2}}
\left(\partial^\rho Z_\rho W_\mu+2 Z^\rho \partial_\rho W_\mu \right)
+\frac{g g^{\prime}}{\sqrt{2\left(g^2+g^{\prime 2}\right)}}\left(\partial^\rho W_\rho A_\mu
+2 W^\rho \partial_\rho A_\mu\right)
+\frac{1}{g}\Xi_\varphi(W,Z,A)\partial_\mu
\nonumber\\ &&
+\frac{i}{\sqrt{2}}\Theta_\varphi(W,W^{\dagger},Z) W_\mu
+\frac{i\left(g^2-g^{\prime 2}\right)}{2g\sqrt{g^2+g^{\prime 2}}}\Xi_\varphi(W,Z,A) Z_\mu
-\frac{ig^{\prime}}{\sqrt{g^2+g^{\prime 2}}}\Xi_\varphi(W^{\dagger},Z,A) A_\mu,
\label{chi_Higgs}
\end{eqnarray}
\begin{eqnarray}
\phi_{\varphi \mu}(W,W^{\dagger},Z,A)&=&
\frac{i}{\sqrt{2}}\left(\partial^\rho \partial_\rho W_\mu^{\dagger}
+2\partial^\rho W_\mu^{\dagger} \partial_\rho
+W_\mu^{\dagger} \partial^\rho \partial_\rho+\partial^\rho W_\rho^{\dagger} \partial_\mu
+2 W^{\dagger \rho} \partial_\rho \partial_\mu\right)
+\frac{\sqrt{g^2+g^{\prime 2}}}{2\sqrt{2}}
\left(\partial^\rho W_\rho^{\dagger} Z_\mu+2 W^{\dagger \rho} \partial_\rho Z_\mu\right)
\nonumber\\&&
-\frac{g^{2}-g^{\prime 2}}{2\sqrt{2\left(g^2+g^{\prime 2}\right)}}
\left(\partial^\rho Z_\rho W_\mu^{\dagger}
+2Z^\rho \partial_\rho W_\mu^{\dagger}\right)
+\frac{g g^{\prime}}{\sqrt{2\left(g^2+g^{\prime 2}\right)}}
\left(\partial^\rho A_\rho W_\mu^{\dagger}+2 A^{\rho}\partial_\rho W_\mu^{\dagger}\right)
\nonumber\\ &&
+\frac{1}{g}\Phi_\varphi(W^{\dagger},Z,A)\partial_\mu
+\frac{i}{\sqrt{2}}\Lambda_\varphi(W,W^{\dagger},Z,A) W_\mu^{\dagger}
-\frac{i\sqrt{g^{2}+g^{\prime 2}}}{2g}\Phi_\varphi(W^{\dagger},Z,A)Z_\mu
\label{phi_Higgs}
\end{eqnarray}
and $\Theta_\varphi(W,W^{\dagger},Z)$, $\Lambda_\varphi(W,W^{\dagger},Z,A)$, $\Xi_\varphi(W,Z,A)$ and
$\Phi_\varphi(W^{\dagger},Z,A)$ are defined as follows:

\begin{eqnarray}
\Theta_\varphi(W,W^{\dagger},Z)&=&-\left[\frac{g^2}{2}\left(W^{\rho} W_\rho^{\dagger} \right)
+\frac{g^2+g^{\prime 2}}{4}\left(Z^\rho Z_\rho\right)\right],
\nonumber\\
\Lambda_\varphi(W,W^{\dagger},Z,A)&=&-\left[\frac{g^2}{2}\left(W^{\dagger \rho} W_\rho\right)
+\frac{\left(g^2-g^{\prime 2}\right)^2}{4\left(g^2+g^{\prime 2}\right)}\left(Z^\rho Z_\rho\right)
+\frac{g^2 g^{\prime 2}}{g^2+g^{\prime 2}}\left(A^\rho A_\rho\right)
-\frac{gg^{\prime}\left(g^{2}-g^{\prime 2}\right)}{g^2+g^{\prime 2}}\left(Z^\rho A_\rho\right)\right],
\nonumber\\
\Xi_\varphi(W,Z,A)&=&-\left[\frac{g\left(g^2-g^{\prime 2}\right)}{2\sqrt{2\left(g^2+g^{\prime 2}
\right)}}\left(W^\rho Z_\rho\right)
-\frac{g\sqrt{g^{2}+g^{\prime 2}}}{2\sqrt{2}}\left(Z^\rho W_{\rho}\right) 
-\frac{g^2 g^{\prime}}{\sqrt{2\left(g^2+g^{\prime 2}\right)}}\left(W^\rho A_\rho\right)\right],
\nonumber\\
\Phi_\varphi(W^{\dagger},Z,A)&=&-\left[-\frac{g\sqrt{g^{2}+g^{\prime 2}}}{2\sqrt{2}}\left(W^{\dagger \rho} Z_\rho\right)
+\frac{g\left(g^2-g^{\prime 2}\right)}{2\sqrt{2\left(g^2+g^{\prime 2}
\right)}}\left(Z^\rho W_\rho^{\dagger}\right)
-\frac{g^2 g^{\prime}}{\sqrt{2\left(g^2+g^{\prime 2}\right)}}
\left(A^{\rho} W_\rho^{\dagger}\right)\right].
\end{eqnarray}
The generalized Higgs Lagrangian is of the following shape:

\begin{eqnarray}
\mathcal{L}_{Higgs}^{\prime}&\equiv&\mathcal{L}_{Higgs}+\mathcal{L}_{Higgs\ \beta}
=\frac{1}{2}\left(\mathcal{D}^\mu \varphi\right)^\dagger \mathcal{D}_\mu\varphi
-\frac{\mu^2}{2}\varphi^{\dagger}\varphi+\frac{\lambda}{4}\left(\varphi^{\dagger}\varphi\right)^2\nonumber\\
&=&\frac{1}{2}\left(D^\mu \varphi \right)^{\dagger} D_\mu \varphi-\frac{\beta}{2}\left(D^\mu \varphi\right)^{\dagger}D^\rho
D_\rho D_\mu \varphi-\frac{\beta}{2}\left(D^\rho D_\rho D^\mu \varphi\right)^{\dagger} D_\mu \varphi
-\frac{\mu^2}{2}\varphi^{\dagger}\varphi+\frac{\lambda}{4}\left(\varphi^{\dagger}\varphi\right)^2,
\label{generalized_Higgs_Lagrangian}
\end{eqnarray}
where $\mathcal{D}_\mu$ is now the generalized covariant derivative 
($\ref{covariant_derivative_electroweak_extended_Higgs}$) as special manifestation of ($\ref{definition_generalized_covariant_derivative}$),
which is based on the representation of the generators of the electroweak theory with respect to the Higgs sector according to
($\ref{generators_phi}$). If the Higgs doublet is expressed as follows:

\begin{equation}
\varphi=\left(\begin{matrix}\varphi^{+}\\ \varphi^{0} \end{matrix}\right),
\end{equation}
the usual Higgs Lagrangian reads  

\begin{eqnarray}
\mathcal{L}_{Higgs}&=&\frac{1}{2}\left[\left(\partial^\mu \varphi^{+}\right)^\dagger \left(\partial_\mu \varphi^{+}\right)
+\left(\partial^\mu \varphi^{0}\right)^\dagger \left(\partial_\mu \varphi^{0}\right)
+\frac{g^2}{2}\varphi^{+ \dagger}\varphi^{+} W^{\mu \dagger}W_\mu
+\frac{g^2}{2}\varphi^{0 \dagger}\varphi^0W^{\mu\dagger} W_\mu
+\frac{\left(g^{2}-g^{\prime 2}\right)^2}{4\left(g^2+g^{\prime 2}\right)}\varphi^{+ \dagger}\varphi^{+} Z^\mu Z_\mu
\right.\nonumber\\&&\left.
+\frac{\left(g^2+g^{\prime 2}\right)}{4}\varphi^{0 \dagger}\varphi^{0} Z^\mu Z_\mu
+\frac{g^2 g^{\prime 2}}{g^2+g^{\prime 2}}\varphi^{+ \dagger} \varphi^{+} A^\mu A_\mu
+\frac{ig}{\sqrt{2}}\left(\partial^\mu \varphi^{+}\right)^{\dagger}\varphi^{0} W_\mu^{\dagger}
+\frac{i\left(g^2-g^{\prime 2}\right)}{2\sqrt{g^2+g^{\prime 2}}}\left(\partial^\mu \varphi^{+}\right)^{\dagger}\varphi^{+} Z_\mu
\right.\nonumber\\&&\left.
-\frac{igg^{\prime}}{\sqrt{g^2+g^{\prime 2}}}\left(\partial^\mu \varphi^{+}\right)^{\dagger}\varphi^{+} A_\mu
+\frac{ig}{\sqrt{2}}\left(\partial^\mu \varphi^{0}\right)^{\dagger}\varphi^{+} W_\mu
-\frac{i\sqrt{g^2+g^{\prime 2}}}{2}\left(\partial^\mu \varphi^{0}\right)^{\dagger}\varphi^{0} Z_\mu
-\frac{ig}{\sqrt{2}}\varphi^{+ \dagger} \left(\partial^\mu \varphi^{0}\right) W_{\mu}^{\dagger}
\right.\nonumber\\&&\left.
-\frac{g\sqrt{g^2+g^{\prime 2}}}{2\sqrt{2}}\varphi^{+ \dagger}\varphi^{0} W^{\mu \dagger}Z_\mu
-\frac{ig}{\sqrt{2}}\varphi^{0 \dagger}\left(\partial^\mu \varphi^{+}\right)W_\mu
+\frac{g\left(g^2-g^{\prime 2}\right)}{2\sqrt{2\left(g^2+g^{\prime 2}\right)}}\varphi^{0 \dagger}\varphi^{+} W^\mu Z_\mu
\right.\nonumber\\&&\left.
-\frac{g^2 g^{\prime}}{\sqrt{2\left(g^2+g^{\prime 2}\right)}}\varphi^{0 \dagger}\varphi^{+} W^{\mu}A_\mu
-\frac{i\left(g^2-g^{\prime 2}\right)}{2\sqrt{g^2+g^{\prime 2}}}\varphi^{+ \dagger}\left(\partial^\mu \varphi^{+}\right)Z_\mu
+\frac{g\left(g^2-g^{\prime 2}\right)}{2\sqrt{2\left(g^2+g^{\prime 2}\right)}}\varphi^{+\dagger}\varphi^{0} W^{\mu \dagger}Z_\mu
\right.\nonumber\\&&\left.
-\frac{gg^{\prime}\left(g^{2}-g^{\prime 2}\right)}{2\left(g^2+g^{\prime 2}\right)}\varphi^{+ \dagger}\varphi^{+} Z^\mu A_\mu
+\frac{i\sqrt{g^2+g^{\prime 2}}}{2}\varphi^{0 \dagger}\left(\partial^\mu \varphi^{0}\right)Z_\mu
-\frac{g\sqrt{g^2+g^{\prime 2}}}{2\sqrt{2}}\varphi^{0 \dagger}\varphi^{+} Z^\mu W_\mu
\right.\nonumber\\&&\left.
+\frac{igg^{\prime}}{\sqrt{g^2+g^{\prime 2}}}\varphi^{+ \dagger} \left(\partial^\mu \varphi^{+}\right)A_\mu
-\frac{g^2 g^{\prime}}{\sqrt{2\left(g^2+g^{\prime 2}\right)}}\varphi^{+ \dagger}\varphi^{0} A^\mu W_\mu^{\dagger}
-\frac{gg^{\prime}\left(g^{2}-g^{\prime 2}\right)}{2\left(g^2+g^{\prime 2}\right)}\varphi^{+ \dagger}\varphi^{+} A^\mu Z_\mu\right]
-\frac{\mu^2}{2}\varphi^{\dagger}\varphi+\frac{\lambda}{4}\left(\varphi^{\dagger}\varphi\right)^2\nonumber\\
\label{usual_Higgs_Lagrangian}
\end{eqnarray}
and the extension of the Higgs Lagrangian reads

\begin{eqnarray}
\mathcal{L}_{Higgs\ \beta}&=&-\frac{\beta}{2}\left\{
\left(\partial^\mu \varphi^{+}\right)^{\dagger}\left(\partial^\rho \partial_\rho \partial_\mu \varphi^{+}\right)
+\left(\partial^\mu \varphi^{0}\right)^{\dagger}\left(\partial^\rho \partial_\rho \partial_\mu \varphi^{0}\right)
-\frac{ig}{\sqrt{2}}\varphi^{+ \dagger}\left(\partial^\rho \partial_\rho \partial^\mu \varphi^{0}\right)W_\mu^{\dagger}
-\frac{ig}{\sqrt{2}}\varphi^{0 \dagger}\left(\partial^\rho \partial_\rho \partial^\mu \varphi^{+}\right)W_\mu
\right.\nonumber\\&&\left.
-\frac{i\left(g^2-g^{\prime 2}\right)}{2\sqrt{g^2+g^{\prime 2}}}\varphi^{+ \dagger}\left(\partial^\rho \partial_\rho \partial^\mu \varphi^{+}\right)Z_\mu
+\frac{i\sqrt{g^2+g^{\prime 2}}}{2}\varphi^{0 \dagger}\left(\partial^\rho \partial_\rho \partial^\mu \varphi^{0}\right)Z_\mu
+\frac{igg^{\prime}}{\sqrt{g^2+g^{\prime 2}}}\varphi^{+ \dagger}\left(\partial^\rho \partial_\rho \partial^\mu \varphi^{0}\right)A_\mu
\right.\nonumber\\&&\left.
+\left(\partial^\rho \partial_\rho \partial^\mu \varphi^{+}\right)^{\dagger}\left(\partial_\mu \varphi^{+}\right)
+\left(\partial^\rho \partial_\rho \partial^\mu \varphi^{0}\right)^{\dagger}\left(\partial_\mu \varphi^{0}\right)
+\frac{ig}{\sqrt{2}}\left(\partial^\rho \partial_\rho \partial^\mu \varphi^{0}\right)^{\dagger}\varphi^{+} W_\mu
+\frac{ig}{\sqrt{2}}\left(\partial^\rho \partial_\rho \partial^\mu \varphi^{+}\right)^{\dagger}\varphi^{0} W_\mu^{\dagger}
\right.\nonumber\\&&\left.
+\frac{i\left(g^2-g^{\prime 2}\right)}{2\sqrt{g^2+g^{\prime 2}}}\left(\partial^\rho \partial_\rho \partial^\mu \varphi^{+}\right)^{\dagger}\varphi^{+} Z_\mu
-\frac{i\sqrt{g^2+g^{\prime 2}}}{2}\left(\partial^\rho \partial_\rho \partial^\mu \varphi^{0}\right)^{\dagger}\varphi^{0} Z_\mu
-\frac{igg^{\prime}}{\sqrt{g^2+g^{\prime 2}}}\left(\partial^\rho \partial_\rho \partial^\mu \varphi^{0}\right)^{\dagger}\varphi^{+} A_\mu
\right.\nonumber\\&&\left.
+\left(\partial^\mu \varphi^{+}\right)^{\dagger}\left[\lambda_{\varphi \mu}(W,W^{\dagger},Z,A) \varphi^{+}\right]
+\left(\partial^\mu \varphi^{+}\right)^{\dagger}\left[\phi_{\varphi \mu}(W,W^{\dagger},Z,A) \varphi^{0}\right]
+\left(\partial^\mu \varphi^{0}\right)^{\dagger}\left[\theta_{\varphi \mu}(W,W^{\dagger},Z,A) \varphi^{0}\right]
\right.\nonumber\\&&\left.
+\left(\partial^\mu \varphi^{0}\right)^{\dagger}\left[\chi_{\varphi \mu}(W,W^{\dagger},Z,A) \varphi^{+}\right]
-\frac{ig}{\sqrt{2}}\varphi^{+ \dagger}\left[\theta_\varphi^\mu(W,W^{\dagger},Z,A) \varphi^{0}\right]W_{\mu}^{\dagger}
-\frac{ig}{\sqrt{2}}\varphi^{+ \dagger}\left[\chi_\varphi^\mu(W,W^{\dagger},Z,A) \varphi^{+}\right]W_{\mu}^{\dagger}
\right.\nonumber\\&&\left.
-\frac{ig}{\sqrt{2}}\varphi^{0 \dagger}\left[\lambda_\varphi^\mu(W,W^{\dagger},Z,A) \varphi^{+}\right]W_{\mu}
-\frac{ig}{\sqrt{2}}\varphi^{0 \dagger}\left[\phi_\varphi^\mu(W,W^{\dagger},Z,A) \varphi^{0}\right]W_{\mu}
-\frac{i\left(g^{2}-g^{\prime 2}\right)}{2\sqrt{g^2+g^{\prime2}}}\varphi^{+\dagger}
\left[\lambda_\varphi^\mu(W,W^{\dagger},Z,A) \varphi^{+}\right]Z_\mu 
\right.\nonumber\\&&\left.
-\frac{i\left(g^{2}-g^{\prime 2}\right)}{2\sqrt{g^2+g^{\prime 2}}}\varphi^{+\dagger}
\left[\phi_\varphi^\mu(W,W^{\dagger},Z,A) \varphi^{0}\right]Z_\mu
+\frac{i\sqrt{g^2+g^{\prime 2}}}{2} Z^\mu \varphi^{0 \dagger}\left[\theta_{\varphi \mu}(W,W^{\dagger},Z,A)\varphi^{0}\right]
\right.\nonumber\\&&\left.
+\frac{i\sqrt{g^2+g^{\prime 2}}}{2} Z^\mu \varphi^{0 \dagger}\left[\chi_{\varphi \mu}(W,W^{\dagger},Z,A) \varphi^{+}\right]
+\frac{igg^{\prime}}{\sqrt{g^2+g^{\prime 2}}}\varphi^{+ \dagger}
\left[\lambda_\varphi^\mu(W,W^{\dagger},Z,A) \varphi^{+}\right]A_\mu
\right.\nonumber\\&&\left.
+\frac{igg^{\prime}}{\sqrt{g^2+g^{\prime 2}}}\varphi^{+ \dagger}
\left[\phi_\varphi^\mu(W,W^{\dagger},Z,A) \varphi^{0}\right]A_\mu
+\left[\lambda_\varphi^\mu(W,W^{\dagger},Z,A) \varphi^{+}\right]^{\dagger}\left(\partial_\mu \varphi^{+}\right)
+\left[\phi_\varphi^\mu(W,W^{\dagger},Z,A) \varphi^{0}\right]^{\dagger}\left(\partial_\mu \varphi^{+}\right)
\right.\nonumber\\&&\left.
+\left[\theta_\varphi^\mu(W,W^{\dagger},Z,A) \varphi^{0}\right]^{\dagger}\left(\partial_\mu \varphi^{0}\right)
+\left[\chi_\varphi^\mu(W,W^{\dagger},Z,A) \varphi^{+}\right]^{\dagger}\left(\partial_\mu \varphi^{0}\right)
+\frac{ig}{\sqrt{2}}\left[\theta_\varphi^\mu(W,W^{\dagger},Z,A) \varphi^{0}\right]^{\dagger}\varphi^{+} W_\mu 
\right.\nonumber\\&&\left.
+\frac{ig}{\sqrt{2}}\left[\chi_\varphi^\mu(W,W^{\dagger},Z,A) \varphi^{+} \right]^{\dagger}\varphi^{+} W_\mu
+\frac{ig}{\sqrt{2}}\left[\lambda_\varphi^\mu(W,W^{\dagger},Z,A) \varphi^{+} \right]^{\dagger}\varphi^{0} W_\mu^{\dagger}
+\frac{ig}{\sqrt{2}}\left[\phi_\varphi^\mu(W,W^{\dagger},Z,A) \varphi^{0} \right]^{\dagger}\varphi^{0} W_\mu^{\dagger}
\right.\nonumber\\&&\left.
+\frac{i\left(g^{2}-g^{\prime 2}\right)}{2\sqrt{g^2+g^{\prime 2}}}\left[\lambda_\varphi^\mu(W,W^{\dagger},Z,A) \varphi^{+}\right]^{\dagger}\varphi^{+}Z_\mu 
+\frac{i\left(g^{2}-g^{\prime 2}\right)}{2\sqrt{g^2+g^{\prime 2}}}\left[\phi_\varphi^\mu(W,W^{\dagger},Z,A) \varphi^{0}\right]^{\dagger}\varphi^{+}Z_\mu
\right.\nonumber\\&&\left.
-\frac{i\sqrt{g^2+g^{\prime 2}}}{2} Z^\mu \left[\theta_{\varphi \mu}(W,W^{\dagger},Z,A) \varphi^{0}\right]^{\dagger}\varphi^{0}
-\frac{i\sqrt{g^2+g^{\prime 2}}}{2} Z^\mu \left[\chi_{\varphi \mu}(W,W^{\dagger},Z,A) \varphi^{+}\right]^{\dagger}\varphi^{0}
\right.\nonumber\\&&\left.
-\frac{igg^{\prime}}{\sqrt{g^2+g^{\prime 2}}}\left[\lambda_\varphi^\mu(W,W^{\dagger},Z,A) \varphi^{+}\right]^{\dagger}\varphi^{+} A_\mu 
-\frac{igg^{\prime}}{\sqrt{g^2+g^{\prime 2}}}\left[\phi_\varphi^\mu(W,W^{\dagger},Z,A) \varphi^{0}\right]^{\dagger}\varphi^{+}A_\mu
\right\},
\label{Higgs_Lagrangian_beta}
\end{eqnarray}
where $\theta_{\varphi \mu}(W,W^{\dagger},Z,A)$, $\lambda_{\varphi \mu}(W,W^{\dagger},Z,A)$,
$\chi_{\varphi \mu}(W,W^{\dagger},Z,A)$ and $\phi_{\varphi \mu}(W,W^{\dagger},Z,A)$ are defined according to
($\ref{theta_Higgs}$),($\ref{lambda_Higgs}$),($\ref{chi_Higgs}$) and ($\ref{phi_Higgs}$).
($\ref{generalized_Higgs_Lagrangian}$) with ($\ref{usual_Higgs_Lagrangian}$) and ($\ref{Higgs_Lagrangian_beta}$)
contains the generalized uncertainty principle and is invariant under local gauge transformations of the form

\begin{eqnarray}
\left(\begin{matrix} \varphi^{+}\\ \varphi_0 \end{matrix}\right)&\rightarrow&
\exp\left[i\alpha_a(x) \tau^a+i\alpha_y(x) y\right]
\left(\begin{matrix} \varphi^{+} \\ \varphi_0 \end{matrix}\right)
\nonumber\\ \mathcal{D}_\mu &\rightarrow& \exp\left[i\alpha_a(x) \tau^a+i\alpha_y(x) y\right]\mathcal{D}_\mu
\exp\left[-i\alpha_a(x) \tau^a-i\alpha_y(x) y\right]
\quad,\quad a=1...3.
\end{eqnarray}
Because of the potential the scalar field $\varphi$ has a vacuum expectation value $\langle \varphi \rangle=\sqrt{\frac{\mu^2}{\lambda}}\equiv \nu$.
The gauge can be chosen in such a way that the component $\varphi^{+}$ vanishes which means that the vacuum expectation values of
the components of the Higgs field, $\varphi^{+}$ and $\varphi^{0}$, are given by

\begin{equation}
\langle \varphi^{+} \rangle=0\quad,\quad \langle \varphi^{0} \rangle=\nu
\end{equation}
and the expansion around the vacuum expectation value looks as follows:

\begin{equation}
\varphi=\left(\begin{matrix}0\\\nu+H\end{matrix}\right).
\end{equation}
Since all terms of the extended Lagrangian of the Higgs field contain additional derivative or potential
factors with respect to the original terms, the generated mass terms for the W and Z-bosons do not differ
from the usual ones and therefore the masses of the W- and Z-bosons remain the same

\begin{equation}
m_W=\frac{vg}{2} \quad,\quad m_Z=\frac{\nu \sqrt{g^2+g^{\prime 2}}}{2}.
\end{equation}
The same holds for the electron mass arising from the Yukawa coupling term

\begin{equation}
\mathcal{L}_{Yukawa}^{\prime}=-G_e\left(\nu_{e L}, e_L\right)\left(\begin{matrix}\varphi^{+}\\\varphi^{0}\end{matrix}\right)e_R+h.c.,
\end{equation}
which is of course not changed as already mentioned above and yields the electron mass

\begin{equation}
m_e=G_e \nu.
\end{equation}

\subsection{Gauge Field Sector of the Generalized Electroweak Theory}

The Lagrangian of the gauge gauge field sector is a special manifestation of the general case of the generalized Yang-Mills Lagrangian ($\ref{generalized_Yang-Mills_Lagrangian}$)
with ($\ref{usual_Yang-Mills_Lagrangian}$),($\ref{Lagrangian_A2}$),($\ref{Lagrangian_A3}$),($\ref{Lagrangian_A4}$),($\ref{Lagrangian_A5}$) and ($\ref{Lagrangian_A6}$).
In the special case of the generalized electroweak theory there are four generators. If the three generators of the $SU(2)$ with respect to the weak isospin and the
hypercharge are related to the general generators of ($\ref{generalized_Yang-Mills_Lagrangian}$) in the following way

\begin{eqnarray}
T^0=\frac{1}{2}{\bf 1} \quad,\quad T^1=\frac{1}{2}\sigma^1 \quad,\quad
T^2=\frac{1}{2}\sigma^2 \quad,\quad T^3=\frac{1}{2}\sigma^3,
\end{eqnarray}
this means for the corresponding structure constants

\begin{equation}
f^{ijk}=\begin{cases}\epsilon^{ijk}\ {\rm for}\ i,j,k \in \{1...3\},\ {\rm where}\ \epsilon^{ijk}\ {\rm denotes\ the\ total\ antisymmetric\ tensor}\\
0,\ {\rm if\ at\ least\ one\ of\ the\ indices\ is\ equal\ to\ zero},
\end{cases}
\label{structure_constants_electroweak}
\end{equation}
since $[T^i,T^j]=i\epsilon^{ijk} T^k$, $[T^i,T^0]=0$ and $[T^0,T^0]=0$ for $i,j,k \in \{1...3\}$.
To determine the corresponding components of the vector field, which are assumed to contain the gauge coupling constants
with respect to this consideration, there have to be used the relations ($\ref{definition_WW*ZA}$) which define the
vector fields $W^\mu$, $W^{\dagger \mu}$, $Z^\mu$ and $A^\mu$ leading to

\begin{equation}
A_0^\mu=\frac{g^{\prime 2} Z^\mu}{\sqrt{g^2+g^{\prime 2}}}+\frac{gg^{\prime} A_\mu}{\sqrt{g^2+g^{\prime 2}}},\quad
A_1^\mu=\frac{g}{\sqrt{2}}\left(W^\mu+W^{\dagger \mu}\right),\quad
A_2^\mu=\frac{g}{\sqrt{2}}\left(W^\mu-W^{\dagger \mu}\right),\quad
A_3^\mu=\frac{g^2 Z^\mu}{\sqrt{g^2+g^{\prime 2}}}-\frac{gg^{\prime} A^\mu}{\sqrt{g^2+g^{\prime 2}}},
\label{vectorfield_components_WW*ZA}
\end{equation}
if the components of the gauge field are modified by the corresponding gauge coupling constants as additional factors.
This means that the gauge field sector of the electroweak theory is the Lagrangian
($\ref{generalized_Yang-Mills_Lagrangian}$) with ($\ref{usual_Yang-Mills_Lagrangian}$),($\ref{Lagrangian_A2}$),
($\ref{Lagrangian_A3}$),($\ref{Lagrangian_A4}$),($\ref{Lagrangian_A5}$) and ($\ref{Lagrangian_A6}$), if one assumes that
the index runs from $0$ to $3$ and the structure constants and vector field components with respect to the Lie Algebra
space correspond to ($\ref{structure_constants_electroweak}$) and ($\ref{vectorfield_components_WW*ZA}$). This Lagrangian
contains the generalized uncertainty principle and is invariant under local gauge transformations of the form

\begin{equation}
\mathcal{D}_\mu \rightarrow \exp\left[i\alpha^a(x) T^a\right]\mathcal{D}_\mu \exp\left[-i\alpha^a(x)T^a\right]
\quad,\quad a =0...3.
\end{equation}

\section{Relation to Quantum Chromodynamics in the Quark Sector}

The Lagrangian of quantum chromodynamics itself under incorporation of a generalized uncertainty principle can obviously be
obtained very easily, if there is known the above consideration of the general case of Yang-Mills theories. One has just to
consider the special case of the $SU(3)_c$ in the above consideration of general Yang-Mills theories and then the corresponding
special Lagrangians ($\ref{generalized_matter_Lagrangian_Yang-Mills}$) and ($\ref{generalized_Yang-Mills_Lagrangian}$) describe
the theory, if as generators $T^a$ the eight Gell-Mann matrices are used and the $f^{abc}$ are accordingly identified with their
structure constants. However, things become more intricate, if there is explored the combination of the electroweak theory and
quantum chromodynamics which means that there is considered the full gauge group of the standard model,
$SU(3)_c \otimes SU(2_L) \otimes U_Y(1)$,
as it appears in the quark sector. This is the case because of the fact that a combination of various gauge groups leads to
a coupling between the different potentials corresponding to these gauge groups, if there is presupposed a generalized uncertainty
principle. This is already the case within the generalized lepton Lagrangian ($\ref{generalized_lepton_Lagrangian}$) and also
within the extended electroweak gauge field Lagrangian, where the photon field couples to the W- and Z-bosons. But in the quark
sector even appears a combination of the strong and the electroweak interaction which are completely independent of each
other in the usual formulation of the standard model. The quark sector of the standard model contains three douplets with
respect to the $SU(2)_L \otimes U(1)_Y$

\begin{equation}
\left(\frac{1+\gamma_5}{2}\right)\left(\begin{matrix}u\\ a_u \end{matrix}\right)\quad,\quad \left(\frac{1+\gamma_5}{2}\right)
\left(\begin{matrix}c\\ a_c \end{matrix}\right)\quad,\quad \left(\frac{1+\gamma_5}{2}\right)\left(\begin{matrix}t\\ a_t \end{matrix}\right),
\label{quark_doublets}
\end{equation}
where $u$, $c$ and $t$ denote the components describing the up-,charm and top-quark and the components $a_u$, $a_c$ and $a_t$
are defined as follows:

\begin{equation}
\left(\begin{matrix}a_u\\a_c\\a_t\end{matrix}\right)=\left(\begin{matrix}V_{ud} & V_{us} & V_{ub}\\V_{cd} & V_{cs} & V_{cb}
\\V_{td} & V_{ts} & V_{tb}\end{matrix}\right)\left(\begin{matrix} d\\s\\b \end{matrix}\right),
\end{equation}
where $d$, $s$ and $b$ denote the states of the down-, strange- and bottom-quark and $V$ denotes the
Cabibbo-Kobayashi-Maskawa-matrix, which generates a mixture of the quarks which third component of
the weak isospin is equal to -1/2. The corresponding covariant derivative is

\begin{equation}
D_\mu=\partial_\mu+iA_\mu^a \tau^a-\frac{i}{3}B_\mu y+i\mathcal{G}_{C} C_\mu^c G^c \quad,\quad a=1...3 \quad,\quad c=1...8,
\label{covariant_derivative_quarks}
\end{equation}
where $\tau^a$ and $y$ are defined according to ($\ref{generators_electroweak_fermion}$),
the $C_\mu^a$ are the components of the gauge potential belonging to the $SU(3)_c$, correspondingly
the $G^a$ denote the eight Gell-Mann matrices with respect to the colour space as generators of the $SU(3)_c$
and $\mathcal{G}_{C}$ describes the coupling constant of the strong interaction. The different factor of
the $y$-term arises from the different hypercharge of the quarks leading according to
($\ref{charge_isospin_hypercharge}$) to the charge values of $q=+2/3$ and $q=-1/3$ respectively.
Using the following definitions:

\begin{eqnarray}
\mathcal{A}_\mu&=&\cos \theta_W A_\mu^a \left(\frac{1+\gamma_5}{4}\right)\sigma^a
-\frac{\sin \theta_W}{3} B_\mu \left[\left(\frac{1+\gamma_5}{4}\right){\bf 1}
+\left(\frac{1-\gamma_5}{2}\right)\right],\quad
\mathcal{G}_{E}=\sqrt{g^2+g^{\prime 2}},\quad \mathcal{C}_\mu=C_\mu^a G^a,
\label{complete_potential}
\end{eqnarray}
the covariant derivative ($\ref{covariant_derivative_quarks}$) reads

\begin{equation}
D_\mu=\partial_\mu+i\mathcal{G}_{E}\mathcal{A}_\mu+i\mathcal{G}_{C}\mathcal{C}_\mu.
\label{covariant_derivative_combined}
\end{equation}
If this expression ($\ref{covariant_derivative_combined}$) of the covariant derivative is inserted into the general expression
for the generalized covariant derivative ($\ref{definition_generalized_covariant_derivative}$), one obtains

\begin{eqnarray}
\mathcal{D}_\mu&=&\left[1-\beta\left(\partial^\rho+i\mathcal{G}_{E}\mathcal{A}^\rho+i\mathcal{G}_{C}\mathcal{C}^\rho\right)
\left(\partial_\rho+i\mathcal{G}_{E}\mathcal{A}_\rho+i\mathcal{G}_{C}\mathcal{C}_\rho\right)\right]
\left(\partial_\mu+i\mathcal{G}_{E}\mathcal{A}_\mu+i\mathcal{G}_{C}\mathcal{C}_\mu\right)\nonumber\\
&=&\partial_\mu+i\mathcal{G}_{E}\mathcal{A}_\mu+i\mathcal{G}_{C}\mathcal{C}_\mu
-\beta\left(
\partial^\rho \partial_\rho \partial_\mu
+i\mathcal{G}_{E}\partial^\rho \partial_\rho \mathcal{A}_\mu
+2i\mathcal{G}_{E}\partial^\rho \mathcal{A}_\mu \partial_\rho
+i\mathcal{G}_{E}\mathcal{A}_\mu \partial^\rho \partial_\rho
+i\mathcal{G}_{C}\partial^\rho \partial_\rho \mathcal{C}_\mu
+2i\mathcal{G}_{C}\partial^\rho \mathcal{C}_\mu \partial_\rho
\right.\nonumber\\&&\left.
+i\mathcal{G}_{C}\mathcal{C}_\mu \partial^\rho \partial_\rho
+i\mathcal{G}_{E}\partial^\rho \mathcal{A}_\rho \partial_\mu
-\mathcal{G}_{E}^2\partial^\rho \mathcal{A}_\rho\mathcal{A}_\mu
-\mathcal{G}_{E} \mathcal{G}_{C} \partial^\rho \mathcal{A}_\rho \otimes \mathcal{C}_\mu
+2i\mathcal{G}_{E}\mathcal{A}^\rho \partial_\rho \partial_\mu
-2\mathcal{G}_{E}^2\mathcal{A}^\rho \partial_\rho \mathcal{A}_\mu
-2\mathcal{G}_{E}^2\mathcal{A}^\rho \mathcal{A}_\mu \partial_\rho
\right.\nonumber\\&&\left.
-2\mathcal{G}_{E}\mathcal{G}_{C}\mathcal{A}^\rho \otimes \partial_\rho \mathcal{C}_\mu
-2\mathcal{G}_{E}\mathcal{G}_{C}\mathcal{A}^\rho \otimes \mathcal{C}_\mu \partial_\rho 
+i\mathcal{G}_{C}\partial^\rho \mathcal{C}_\rho \partial_\mu
-\mathcal{G}_{E}\mathcal{G}_{C}\mathcal{A}_\mu \otimes \partial^\rho \mathcal{C}_\rho
-\mathcal{G}_{C}^2 \partial^\rho \mathcal{C}_\rho \mathcal{C}_\mu
+2i\mathcal{G}_{C}\mathcal{C}^\rho \partial_\rho \partial_\mu
\right.\nonumber\\&&\left.
-2\mathcal{G}_{E}\mathcal{G}_{C}\partial^\rho \mathcal{A}_\mu \otimes \mathcal{C}_\rho 
-2\mathcal{G}_{E}\mathcal{G}_{C}\mathcal{A}_\mu \otimes \mathcal{C}^\rho \partial_\rho 
-2\mathcal{G}_{C}^2\mathcal{C}^\rho \partial_\rho \mathcal{C}_\mu
-2\mathcal{G}_{C}^2\mathcal{C}^\rho \mathcal{C}_\mu \partial_\rho
-\mathcal{G}_{E}^2\mathcal{A}^\rho \mathcal{A}_\rho \partial_\mu
-i\mathcal{G}_{E}^3\mathcal{A}^\rho \mathcal{A}_\rho \mathcal{A}_\mu 
\right.\nonumber\\&&\left.
-i\mathcal{G}_{E}^2\mathcal{G}_{C} \mathcal{A}^\rho \mathcal{A}_\rho \otimes \mathcal{C}_\mu
-\mathcal{G}_{C}^2\mathcal{C}^\rho \mathcal{C}_\rho \partial_\mu
-i\mathcal{G}_{E}\mathcal{G}_{C}^2 \mathcal{A}_\mu \otimes \mathcal{C}^\rho \mathcal{C}_\rho
-i\mathcal{G}_{C}^3\mathcal{C}^\rho \mathcal{C}_\rho \mathcal{C}_\mu
-2\mathcal{G}_{E}\mathcal{G}_{C} \mathcal{A}^\rho \otimes \mathcal{C}_\rho \partial_\mu
\right.\nonumber\\&&\left.
-2i\mathcal{G}_{E}^2\mathcal{G}_{C} \mathcal{A}^\rho \mathcal{A}_\mu \otimes \mathcal{C}_\rho 
-2i\mathcal{G}_{E}\mathcal{G}_{C}^2 \mathcal{A}^\rho \otimes \mathcal{C}_\rho \mathcal{C}_\mu
\right).
\label{covariant_derivative_quark_sector}
\end{eqnarray}
Within the covariant derivative ($\ref{covariant_derivative_quark_sector}$) obviously appear terms which contain
products of $\mathcal{A}_\mu$ and $\mathcal{C}_\mu$, which live in the product space of the $SU(3)_c$ Lie algebra
of quantum chromodynamics and  the $SU(2)_L \otimes U(1)_Y$ Lie algebra of the electroweak theory, which
generators are tensor products of the Gell-Mann matrices with the generators of the electroweak theory ($\ref{generators_electroweak_fermion}$).
This means nothing else than that the gluons mediating the strong interaction and the mediation particles of the electroweak
interaction are coupled to each other according to this generalized version of the description of the interactions of particle
physics. This holds for the quark sector as well as for the gauge field sector, which both contain the covariant derivative
($\ref{covariant_derivative_quark_sector}$). If the three left handed quark doublets are subsumed into $Q_L$ and the
flavour triplets of the right handed quarks with charge $q=+2/3$ and $q=-1/3$ are subsumed into $Q_R^{+}$ and $Q_R^{-}$

\begin{equation}
Q_L=\left(\frac{1+\gamma_5}{2}\right)
\left\{\left(\begin{matrix}u\\ a_u \end{matrix}\right),
\left(\begin{matrix}c\\ a_c \end{matrix}\right),
\left(\begin{matrix}t\\ a_t \end{matrix}\right)\right\}
\quad,\quad
Q_R^{+}=\left(\frac{1-\gamma_5}{2}\right)\left(\begin{matrix}u\\c\\t\end{matrix}\right)
\quad,\quad
Q_R^{-}=\left(\frac{1-\gamma_5}{2}\right)\left(\begin{matrix}a_u\\a_c\\a_t\end{matrix}\right),
\end{equation}
where $Q_L$, $Q_R^{+}$ and $Q_R^{-}$ are of course also vectors in the colour space, then the Lagrangian
of the quark sector of the standard model reads

\begin{eqnarray}
\mathcal{L}_{quark}^{\prime}&=&i\bar Q_L \gamma^\mu \mathcal{D}_\mu Q_L
+i\bar Q_R^{+} \gamma^\mu \mathcal{D}_\mu Q_R^{+}
+i\bar Q_R^{-} \gamma^\mu \mathcal{D}_\mu Q_R^{-}\nonumber\\
&\equiv& \mathcal{L}_{q\ EW}+\mathcal{L}_{q\ EW-QCD}+\mathcal{L}_{q\ QCD}.
\label{generalized_quark_Lagrangian}
\end{eqnarray}
This Lagrangian ($\ref{generalized_quark_Lagrangian}$) contains the generalized gauge principle and is invariant under
local gauge transformations of the form

\begin{eqnarray}
Q_L&\rightarrow& \exp\left[i\alpha_a(x) \tau^a+i\alpha_y(x) y+i\alpha_G^b(x) G^b\right]Q_L\quad,\quad
Q_R^{+/-}\rightarrow \exp\left[i\alpha_y(x) y+i\alpha_G^b(x) G^b\right]Q_R^{+/-},
\nonumber\\ \mathcal{D}_\mu &\rightarrow& \exp\left[i\alpha_a(x)\tau^a+i\alpha_y(x) y+i\alpha_G^b(x) G^b\right]
\mathcal{D}_\mu\exp\left[-i\alpha_a(x) \tau^a-i\alpha_y(x) y-i\alpha_G^b(x) G^b\right],\quad a=1...3,\quad b=1...8.\nonumber\\
\end{eqnarray}
The interesting term in ($\ref{generalized_quark_Lagrangian}$) is $\mathcal{L}_{q\ EW-QCD}$ containing the
intersection between the electroweak and the strong interaction which reads

\begin{equation}
\mathcal{L}_{q\ EW-QCD}=i\bar Q_L \gamma^\mu \Delta_\mu(\mathcal{A},\mathcal{C}) Q_L
+i\bar Q_R^{+} \gamma^\mu \Delta_\mu(\mathcal{A},\mathcal{C}) Q_R^{+}
+i\bar Q_R^{-} \gamma^\mu \Delta_\mu(\mathcal{A},\mathcal{C}) Q_R^{-},
\label{Lagrangian_EW-QCD}
\end{equation}
where $\Delta_\mu(\mathcal{A},\mathcal{C})$ is defined as

\begin{eqnarray}
\Delta_\mu(\mathcal{A},\mathcal{C})&=&
-\mathcal{G}_{E} \mathcal{G}_{C} \partial^\rho \mathcal{A}_\rho \otimes \mathcal{C}_\mu
-2\mathcal{G}_{E}\mathcal{G}_{C}\mathcal{A}^\rho \otimes \partial_\rho \mathcal{C}_\mu
-2\mathcal{G}_{E}\mathcal{G}_{C}\mathcal{A}^\rho \otimes \mathcal{C}_\mu \partial_\rho
-\mathcal{G}_{E}\mathcal{G}_{C}\mathcal{A}_\mu \otimes \partial^\rho \mathcal{C}_\rho \nonumber\\&&
-2\mathcal{G}_{E}\mathcal{G}_{C}\partial^\rho \mathcal{A}_\mu \otimes \mathcal{C}_\rho
-2\mathcal{G}_{E}\mathcal{G}_{C}\mathcal{A}_\mu \otimes \mathcal{C}^\rho \partial_\rho
-i\mathcal{G}_{E}^2\mathcal{G}_{C} \mathcal{A}^\rho \mathcal{A}_\rho \otimes \mathcal{C}_\mu
-i\mathcal{G}_{E}\mathcal{G}_{C}^2 \mathcal{A}_\mu \otimes \mathcal{C}^\rho \mathcal{C}_\rho \nonumber\\&&
-2\mathcal{G}_{E}\mathcal{G}_{C} \mathcal{A}^\rho \otimes \mathcal{C}_\rho \partial_\mu
-2i\mathcal{G}_{E}^2\mathcal{G}_{C} \mathcal{A}^\rho \mathcal{A}_\mu \otimes \mathcal{C}_\rho
-2i\mathcal{G}_{E}\mathcal{G}_{C}^2 \mathcal{A}^\rho \otimes \mathcal{C}_\rho \mathcal{C}_\mu.
\label{Quark_Delta}
\end{eqnarray}
Thus it has been emerged that the strong interaction cannot be separated from the electroweak theory anymore
in a formulation of the standard model containing a generalized uncertainty principle. In such a theory there exist
coupling terms and thus vertices between quarks, mediation particles of the electroweak theory and gluons and this
means that the electroweak interaction is explicitly coupled to the strong interaction.
This result is quite interesting because in a fundamental theory one expects that all interactions are unified.
In this sense this result delivers a hint that the introduction of a generalized uncertainty principle to quantum
field theory could indeed yield an appropriate description of some aspects of a fundamental theory including a quantum
theoretical description of gravity as they are reflected in usual quantum field theory. This holds although these additional
coupling terms represent not really a grater unification, since also gravity couples to the mediation particles of the other
interactions and this is not interpreted as a kind of unified description with the other fundamental interactions.
If there is incorporated the gauge group $SU(3)_c$ of quantum chromodynamics, the Lagrangian of the gauge field of course
also contains these intersection between the gauge bosons of the electroweak theory and the gluons, since to built the
field strength tensor of the complete standard model there has naturally to be used the covariant derivative
($\ref{covariant_derivative_quark_sector}$) containing the gluon potential besides the electroweak potential.
This would lead to a combined gauge field Lagrangian

\begin{equation}
\mathcal{L}_{gauge\ EW-QCD}=\frac{1}{4}{\rm tr}\left[\mathcal{F}_{\mu\nu}(\mathcal{A}\mathcal{C})
\mathcal{F}^{\mu\nu}(\mathcal{A},\mathcal{C})\right],
\label{combined_gauge_field_Lagrangian}
\end{equation}
which combines the gauge fields of the electroweak theory and quantum chromodynamics.
But this Lagrangian will not be calculated explicitly in this paper.

\section{Phenomenological Consequences}

The extended electroweak theory containing a minimal length induced by a generalized uncertainty
principle as it is presented in this paper could become phenomenologically very important with
respect to processes at very high energies as they are reached at the LHC. Therefore in this section
are discussed two classes of production respectively scattering processes being predicted by the theory,
one of them refers to the Higgs sector and the other refers to the quark sector. In the Higgs sector
appear new interaction terms implying additional vertices containg three and four vector bosons in the calculation
to the first order in $\beta$ as it is considered in this paper. This leads to new production processes of the Higgs
particle. In the quark sector the completely new intersection between the electroweak and the strong interaction leads
of course to completely new Feynman diagrams and corresponding scattering processes of quarks. Of course, there
appear also important new interaction processes in the gauge field sector and the sector of the quark Lagrangian
just containing interaction terms with the gluons which could become important as well. In the Feynman graphs drawn
below plain lines denote Higgs particles, plain lines with arrows denote quarks, wiggly lines denote photons,
$Z$-particles, $W^{+}$-particles or $W^{-}$-particles and curly lines denote gluons. These Feynman diagrams represent
classes of interactions and thus correspond to various special interactions. Because of the big number of additional
terms in the generalized Lagrangians it is impossible to consider every special vertex separately.
The terms in the gauge field sector of the highest order in the potential, ($\ref{Lagrangian_A5}$) and ($\ref{Lagrangian_A6}$), induce vertices containing five and six gauge bosons:
\\

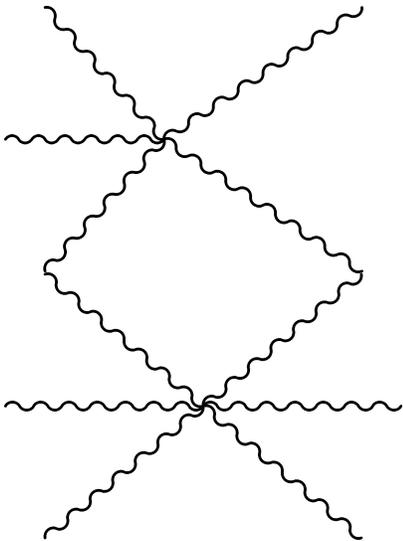
\begin{fmffile}{FeynmanGraph_A}
\begin{fmfgraph*}(150,100)
 \fmfleft{l1,l2,l3}
 \fmf{photon,label=$\gamma,Z,W^{+},W^{-}$}{l1,v}
 \fmf{photon,label=$\gamma,Z,W^{+},W^{-}$}{l2,v}
 \fmf{photon,label=$\gamma,Z,W^{+},W^{-}$}{l3,v}
 \fmf{photon,label=$\gamma,Z,W^{+},W^{-}$}{v,r1}
 \fmf{photon,label=$\gamma,Z,W^{+},W^{-}$}{v,r2}
 \fmfright{r1,r2}
\end{fmfgraph*}
\end{fmffile}

\begin{fmffile}{FeynmanGraph_B}
\begin{fmfgraph*}(150,100)
 \fmfleft{l1,l2,l3}
 \fmf{photon,label=$\gamma,Z,W^{+},W^{-}$}{l1,v}
 \fmf{photon,label=$\gamma,Z,W^{+},W^{-}$}{l2,v}
 \fmf{photon,label=$\gamma,Z,W^{+},W^{-}$}{l3,v}
 \fmf{photon,label=$\gamma,Z,W^{+},W^{-}$}{v,r1}
 \fmf{photon,label=$\gamma,Z,W^{+},W^{-}$}{v,r2}
 \fmf{photon,label=$\gamma,Z,W^{+},W^{-}$}{v,r3}
 \fmfright{r1,r2,r3}
\end{fmfgraph*}
\end{fmffile}
\\
\\
According to the above remark these two classes of vertices describe a big number of special interactions incorporating
photons, $Z$-particles, $W^{+}$-particles and $W^{-}$-particles, which are not mentioned separately here. But they can directly
be obtained from ($\ref{Lagrangian_A5}$) and ($\ref{Lagrangian_A6}$). As all new vertices in this paper, they depend linear on
the parameter $\beta$. In the gauge field sector also takes place an intersection of the electroweak and strong interaction,
leading to a direct coupling of the gauge bosons of the electroweak theory and the gluons of quantum chromodynamics,
which is contained in the Lagrangian ($\ref{combined_gauge_field_Lagrangian}$). In the two following subsections there will be considered new processes induced by the generalization of the Higgs Lagrangian and the intersection Lagrangian of the
generalized quark Lagrangian describing interactions incorporating gluons and electroweak gauge bosons.

\subsection{New Processes Concerning the Production of Higgs Particles}

Since there exists a great probability that the Higgs particle as the only constitutive ingredient of the 
standard model which has not been discovered so far will be discovered at the LHC, the new possible processes
leading to a production of Higgs particles according to the extension of the electroweak theory presented in
this paper are of particular interest. The new possible vertices can be obtained by the extended interaction
terms contained in the extension term of the Lagrangian ($\ref{Higgs_Lagrangian_beta}$). The first class of
new vertices describes interactions between three vector bosons and two Higgs particles and the vertices are
therefore of the following shape:
\\

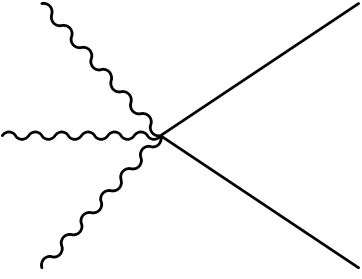
\begin{fmffile}{FeynmanGraph_C}
\begin{fmfgraph*}(150,100)
 \fmfleft{l1,l2,l3} 
 \fmf{photon,label=$\gamma,Z,W^{+},W^{-}$}{l1,v}
 \fmf{photon,label=$\gamma,Z,W^{+},W^{-}$}{l2,v}
 \fmf{photon,label=$\gamma,Z,W^{+},W^{-}$}{l3,v}
 \fmf{plain,label=$H$}{v,r1}
 \fmf{plain,label=$H$}{v,r2}
 \fmfright{r1,r2}
\end{fmfgraph*}
\end{fmffile}
\\
\\
The concrete mathematical expressions belonging to the very big number of special vertices which are contained
in this class of vertices can directly be obtained from the Lagrangian ($\ref{Higgs_Lagrangian_beta}$), but again
are not explicitly written here.
These special vertices correspond to the following processes implying the production of a pair of Higgs
particles: $\gamma \gamma \gamma \rightarrow H H$, $Z Z Z \rightarrow H H$, 
$W^{+} W^{+} W^{-}\rightarrow H H$, $W^{+} W^{-} W^{-} \rightarrow H H$, $\gamma \gamma W^{+}\rightarrow H H$, 
$\gamma \gamma W^{-}\rightarrow H H$, $\gamma \gamma Z \rightarrow H H$, 
$\gamma Z Z \rightarrow H H$, $\gamma W^{+} W^{-}\rightarrow H H$, $\gamma Z W^{-}\rightarrow H H$, 
$\gamma Z W^{+}\rightarrow H H$, $Z W^{+} W^{-}\rightarrow H H$, $Z Z W^{+}\rightarrow H H$, 
$Z Z W^{-}\rightarrow H H$.
Of course, the relevance of the contribution of these processes depends on the scale of the parameter $\beta$
defining the strength of the modification of the uncertainty relation and thus the value of the minimal length.
The corresponding cross sections are proportional to $\beta^2$: $\sigma \sim \beta^2$, and they are of quadratic
dimension in the coupling constants of the electroweak theory, $g$ or $g^{\prime}$ respectively. The second class
of vertices describes interactions between four vector bosons and two Higgs particles. The corresponding special
production processes are not considered here. 
\\

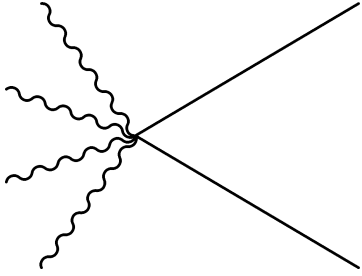
\begin{fmffile}{FeynmanGraph_D}
\begin{fmfgraph*}(150,100)
 \fmfleft{l1,l2,l3,l4} 
 \fmf{photon,label=$\gamma,Z,W^{+},W^{-}$}{l1,v}
 \fmf{photon,label=$\gamma,Z,W^{+},W^{-}$}{l2,v}
 \fmf{photon,label=$\gamma,Z,W^{+},W^{-}$}{l3,v}
 \fmf{photon,label=$\gamma,Z,W^{+},W^{-}$}{l4,v}
 \fmf{plain,label=$g$}{v,r1}
 \fmf{plain,label=$g$}{v,r2}
 \fmfright{r1,r2}
\end{fmfgraph*}
\end{fmffile}
\\
\\

\subsection{Combined Interaction Processes of the Electroweak and Strong Interaction in the Quark Sector}

As it has been shown above in the quark sector of the extended gauge theory, given in ($\ref{Lagrangian_EW-QCD}$)
with ($\ref{Quark_Delta}$), appear new interaction terms containing the gauge potential of the electroweak theory
as well as the gauge potential of quantum chromodynamics. Therefore one is led to new interaction processes
induced by these terms where the strong and the electroweak interaction are combined. This corresponds to the
introduction of the following classes of vertices:\\
\\
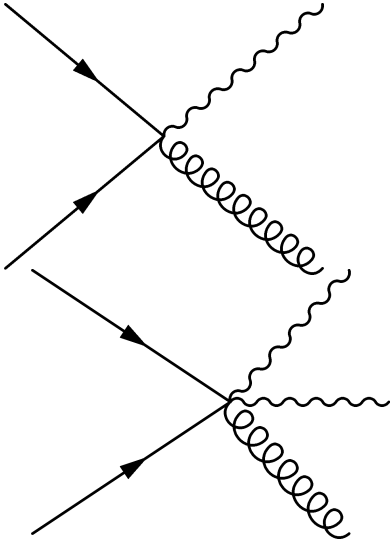
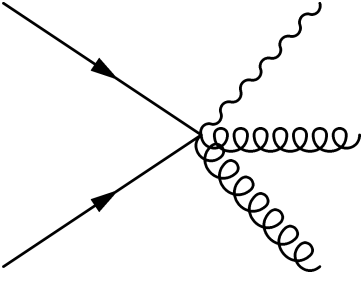
\begin{fmffile}{FeynmanGraph_E}
\begin{fmfgraph*}(150,100)
 \fmfleft{l1,l2}
 \fmf{fermion,label=$q$}{l1,v}
 \fmf{fermion,label=$q$}{l2,v}
 \fmf{curly,label=$g$}{v,r1}
 \fmf{photon,label=$\gamma,Z,W^{+},W^{-}$}{v,r2}
 \fmfright{r1,r2}
\end{fmfgraph*}
\end{fmffile}

\begin{fmffile}{FeynmanGraph_F}
\begin{fmfgraph*}(150,100)
 \fmfleft{l1,l2}
 \fmf{fermion,label=$q$}{l1,v}
 \fmf{fermion,label=$q$}{l2,v}
 \fmf{curly,label=$g$}{v,r1}
 \fmf{photon,label=$\gamma,Z,W^{+},W^{-}$}{v,r2}
 \fmf{photon,label=$\gamma,Z,W^{+},W^{-}$}{v,r3}
 \fmfright{r1,r2,r3}
\end{fmfgraph*}
\end{fmffile}

\begin{fmffile}{FeynmanGraph_G}
\begin{fmfgraph*}(150,100)
 \fmfleft{l1,l2}
 \fmf{fermion,label=$q$}{l1,v}
 \fmf{fermion,label=$q$}{l2,v}
 \fmf{curly,label=$g$}{v,r1}
 \fmf{curly,label=$g$}{v,r2}
 \fmf{photon,label=$\gamma,Z,W^{+},W^{-}$}{v,r3}
 \fmfright{r1,r2,r3}
\end{fmfgraph*}
\end{fmffile}
\\
\\
The first class of vertices describes the interaction of a quark with one gluon and one photon, $Z$-particle,
$W^{+}$-particle or $W^{-}$-particle. The second class of vertices describes the interaction of a quark with
two gauge bosons of the electroweak theory and a gluon and the third class of vertices describes the interaction
of a quark with one gauge boson of the electroweak theory and two gluons. The flavour and colour of the quark
is not determined.

If one considers the scattering process of two quarks at the tree level for example, the Feynman graph below
yields the most important contribution of the intersection term of the extended Lagrangian of the quark
sector, since the two vertices are quadratic in the coupling constant of the strong interaction $\mathcal{G}_C$.
The other two vertices are just linear in the coupling constant of the strong interaction and therefore
the corresponding quark scattering processes give a contribution being much smaller.\\
\\
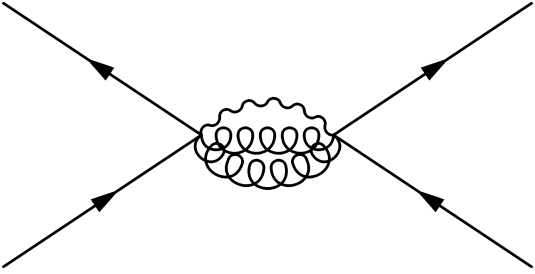
\begin{fmffile}{FeynmanGraph_H}
\begin{fmfgraph*}(250,100)
 \fmfleft{l1,l2}
 \fmf{fermion,label=$q$}{l1,v1,l2}
 \fmf{fermion,label=$q$}{r1,v2,r2}
 \fmf{photon,label=$\gamma,Z,W^{+},W^{-}$,left=0.5}{v1,v2}
 \fmf{curly,label=$g$}{v1,v2}
 \fmf{curly,label=$g$,right=0.5}{v1,v2}
 \fmfright{r1,r2}
\end{fmfgraph*}
\end{fmffile}
\\
\\
Of course, the diagram describing the quark scattering, corresponds to various processes, since
the quark coulour and flavour is not determined and the gauge boson of the electroweak theory
as one of the three mediation particles can be a photon, $Z$-particle, $W^{+}$-particle or
$W^{-}$-particle. Since the two vertices are proportional to $\beta$, $\mathcal{G}_C^2$ and
$\mathcal{G}_E$, the corresponding cross sections of these processes are proportional to
$\beta^4$, $\mathcal{G}_C^8$ and $\mathcal{G}_E^4$: $\sigma \sim \beta^2 \mathcal{G}_C^4 \mathcal{G}_E^2$,
where $\mathcal{G}_E$ is defined according to ($\ref{complete_potential}$). The above diagram
also describes an annihilation respectively production process of quarks mediated by two gluons and one electroweak
gauge boson: $q \bar q \rightarrow gg\gamma/Z/W^{+}/W^{-}\rightarrow q \bar q$. Depending on the special gauge boson
of the electroweak theory, a quark-anti-quark pair can be converted to a quark-anti-quark pair of new flavour. 
It is not possible to explore the rich phenomenology of the extended electroweak theory completely in this paper.
Therefore there has just been given a short discussion of possibly relevant phenomenological consequences arising
from the presented theory without determining concrete expressions for cross sections. But these considerations
should inspire more elaborate investigations of the mentioned phenomenology with respect to continuing
research projects.

\section{Summary and Discussion}

In this paper the extension of gauge theories according to the introduction of a generalized uncertainty principle
to quantum field theory as it was considered in \cite{Kober:2010sj} with respect to electrodynamics and the $SO(3,1)$
gauge formulation of general relativity has been applied to the case of Yang-Mills theories with gauge group $SU(N)$ and
escpecially to the electroweak theory of the standard model with gauge group $SU(2)_L \otimes U(1)_Y$. There have been
calculated the generalized Lagrangians of the matter field being coupled to the Yang-Mills gauge field and the corresponding
Yang-Mills Lagrangian for the gauge field by presupposing the generalized field strength tensor which is built from the
extended covariant derivative. To obtain the correct extension of the specific case of the electroweak theory, the corresponding
covariant derivative containing the $W$-,$W^{\dagger}$-,$Z$- and $A$-field has to be inserted to the general shape of the
generalized covariant derivative containing products of the gauge potential and maintaining gauge invariance in the presence
of a generalized uncertainty principle. There have been calculated the generalized terms of the different sectors of the
corresponding Lagrangian in a series expansion to the first order in the modification parameter $\beta$. This consideration
leads to a much more complicated interaction structure of the electroweak theory with additional self-interaction terms. The
electromagnetic field does not only interact with itself but also with the $W$-, $W^{\dagger}$- and $Z$-field. Although the
dynamics of the Higgs-sector is also changed decisively, the generated masses remain unchanged, since all extension terms
contain at least one additional derivative or potential factor and therefore there appear no further mass terms. The most
interesting property appears in the quark sector, since the consideration of the full gauge group of the standard model,
$SU(3)_c \otimes SU(2)_L \otimes U(1)_Y$, leads to a coupling between the gauge potential of the electroweak theory and the
one of quantum chromodynamics. This quality has its origin in the appearing product of potentials in the generalized covariant
derivative leading to intersections of the potentials, if two gauge groups are combined. Accordingly the introduction of a
generalized uncertainty principle to the description of the standard model causes an interaction between the photon and the
$W$- and $Z$-particles as mediation particles of the electroweak interaction and the gluons as mediation particles of the
strong interaction. This can be interpreted as a kind of consequence of the fact that on a fundamental level the interactions
of the standard model are completely unified and can therefore not be separated from each other anymore. With respect to this
advisement this property of the extended electroweak theory seems not to be very surprising. The presented extension of the
electroweak theory yields further one of the few considerations, like \cite{Chernodub:2007bz} for example, containing an
explicite relation between indirect aspects of quantum gravity and the electroweak theory. Considerations of generalizations
of the standard model with respect to noncommutative geometry can be found in
\cite{Calmet:2001na},\cite{Calmet:2003jv},\cite{Calmet:2006zy}. Of course, the generalization of the electroweak theory in
\cite{Calmet:2001na} is closely related to the generalized description of the electroweak theory presented in this paper, since
the concept of noncommuting coordinates is associated with the generalized uncertainty principle.
Finally there have been considered some special processes which could lead to observable effects at the LHC. This depends
of course on the scale defined by the parameter $\beta$ within the generalized uncertainty principle. The extension
of the Higgs Lagrangian is of particular interest, since the Higgs particle as a decisive ingredient of the standard model is
still undiscovered and is expected to be discovered at the LHC. There appear new processes where the annihilation of three or respectively four gauge bosons leads to the production of two Higgs particles what could give a significant contribution to the production rate of Higgs particles at the LHC. One possible process would be the annihilation of three photons leading to
a production of a pair of Higgs particles, $\gamma \gamma \gamma \rightarrow HH$. Besides, the intersection of the electroweak
and the strong interaction in the quark sector leads to completely new vertices which could give important contributions to quark scattering processes at high energies. The calculation of concrete cross sections with respect to the Higgs as well as the quark sector would be a very interesting topic for further research projects continuing the considerations of this paper.
\\ \noindent
$Acknowledgement$: I would like to thank the Messer Stiftung for financial support.


\begin{thebibliography}{99}

\bibitem{Maggiore:1993kv}
  M.~Maggiore,
  Phys.\ Lett.\  B {\bf 319} (1993) 83
  [arXiv:hep-th/9309034].

\bibitem{Kempf:1994su}
  A.~Kempf, G.~Mangano and R.~B.~Mann,
  Phys.\ Rev.\  D {\bf 52} (1995) 1108
  [arXiv:hep-th/9412167].

\bibitem{Hinrichsen:1995mf}
  H.~Hinrichsen and A.~Kempf,
  J.\ Math.\ Phys.\  {\bf 37} (1996) 2121
  [arXiv:hep-th/9510144].

\bibitem{Kempf:1996ss}
  A.~Kempf,
  J.\ Math.\ Phys.\  {\bf 38} (1997) 1347
  [arXiv:hep-th/9602085].

\bibitem{Kempf:1996nk}
  A.~Kempf and G.~Mangano,
  Phys.\ Rev.\  D {\bf 55} (1997) 7909
  [arXiv:hep-th/9612084].

\bibitem{Lubo:1999xg}
  M.~Lubo,
  Phys.\ Rev.\  D {\bf 61} (2000) 124009
  [arXiv:hep-th/9911191].

\bibitem{Chang:2001kn}
  L.~N.~Chang, D.~Minic, N.~Okamura and T.~Takeuchi,
  Phys.\ Rev.\  D {\bf 65} (2002) 125027
  [arXiv:hep-th/0111181].

\bibitem{Hossenfelder:2003jz}
  S.~Hossenfelder, M.~Bleicher, S.~Hofmann, J.~Ruppert, S.~Scherer and H.~Stoecker,
  Phys.\ Lett.\  B {\bf 575} (2003) 85
  [arXiv:hep-th/0305262].

\bibitem{Harbach:2003qz}
  U.~Harbach, S.~Hossenfelder, M.~Bleicher and H.~Stoecker,
  Phys.\ Lett.\  B {\bf 584} (2004) 109
  [arXiv:hep-ph/0308138].

\bibitem{Harbach:2005yu}
  U.~Harbach and S.~Hossenfelder,
  Phys.\ Lett.\  B {\bf 632} (2006) 379
  [arXiv:hep-th/0502142].

\bibitem{Bleicher:2010dg}
  M.~Bleicher, P.~Nicolini, M.~Sprenger,
  [arXiv:1011.5225 [hep-ph]].

\bibitem{Hossenfelder:2004up}
  S.~Hossenfelder,
  Phys.\ Rev.\  D {\bf 70} (2004) 105003
  [arXiv:hep-ph/0405127].

\bibitem{Hossenfelder:2006cw}
  S.~Hossenfelder,
  Phys.\ Rev.\  D {\bf 73}, 105013 (2006)
  [arXiv:hep-th/0603032].

\bibitem{Hossenfelder:2007re}
  S.~Hossenfelder,
  Class.\ Quant.\ Grav.\  {\bf 25} (2008) 038003
  [arXiv:0712.2811 [hep-th]].

\bibitem{Bang:2006va}
  J.~Y.~Bang and M.~S.~Berger,
  Phys.\ Rev.\  D {\bf 74} (2006) 125012
  [arXiv:gr-qc/0610056].

\bibitem{Buisseret:2010cw}
  F.~Buisseret,
  Phys.\ Rev.\ A{\bf 82 } (2010)  062102.
  [arXiv:1011.3690 [hep-th]].

\bibitem{Nozari:2010qy}
  K.~Nozari, P.~Pedram,
  Europhys.\ Lett.\  {\bf 92 } (2010)  50013.
  [arXiv:1011.5673 [hep-th]].

\bibitem{Shibusa:2007ju}
  Y.~Shibusa,
  Int.\ J.\ Mod.\ Phys.\  A {\bf 22} (2007) 5279
  [arXiv:0704.1545 [hep-th]].

\bibitem{Percacci:2010af}
  R.~Percacci, G.~P.~Vacca,
  Class.\ Quant.\ Grav.\  {\bf 27 } (2010)  245026.
  [arXiv:1008.3621 [hep-th]].

\bibitem{Kim:2008kc}
  Y.~W.~Kim, H.~W.~Lee and Y.~S.~Myung,
  Phys.\ Lett.\  B {\bf 673} (2009) 293
  [arXiv:0811.2279 [gr-qc]].

\bibitem{Matsuo:2005fb}
  T.~Matsuo, Y.~Shibusa,
  Mod.\ Phys.\ Lett.\  {\bf A21 } (2006)  1285-1296.
  [hep-th/0511031].

\bibitem{Moayedi:2010vp}
  S.~K.~Moayedi, M.~R.~Setare, H.~Moayeri,
  Int.\ J.\ Theor.\ Phys.\  {\bf 49 } (2010)  2080-2088.
  [arXiv:1004.0563 [hep-th]].

\bibitem{Maggiore:1993rv}
  M.~Maggiore,
  Phys.\ Lett.\  B {\bf 304} (1993) 65
  [arXiv:hep-th/9301067].

\bibitem{Maggiore:1993zu}
  M.~Maggiore,
  Phys.\ Rev.\  D {\bf 49} (1994) 5182
  [arXiv:hep-th/9305163].

\bibitem{Scardigli:1999jh}
 F.~Scardigli,
 Phys.\ Lett.\  B {\bf 452} (1999) 39
 [arXiv:hep-th/9904025].

\bibitem{Capozziello:1999wx}
  S.~Capozziello, G.~Lambiase and G.~Scarpetta,
  Int.\ J.\ Theor.\ Phys.\  {\bf 39} (2000) 15
  [arXiv:gr-qc/9910017].
  
\bibitem{Crowell:1999up}
  L.~B.~Crowell,
  Found.\ Phys.\ Lett.\  {\bf 12} (1999) 585.

\bibitem{Chang:2001bm}
  L.~N.~Chang, D.~Minic, N.~Okamura and T.~Takeuchi,
  Phys.\ Rev.\  D {\bf 65} (2002) 125028
  [arXiv:hep-th/0201017].

\bibitem{Kim:2006rx}
 W.~Kim, Y.~W.~Kim and Y.~J.~Park,
 Phys.\ Rev.\  D {\bf 74} (2006) 104001
 [arXiv:gr-qc/0605084].

\bibitem{Park:2007az}
 M.~I.~Park,
 Phys.\ Lett.\  B {\bf 659} (2008) 698
 [arXiv:0709.2307 [hep-th]].

\bibitem{Scardigli:2007bw}
  F.~Scardigli and R.~Casadio,
  Int.\ J.\ Mod.\ Phys.\  D {\bf 18} (2009) 319
  [arXiv:0711.3661 [hep-th]].

\bibitem{Kim:2007hf}
 W.~Kim, E.~J.~Son and M.~Yoon,
 JHEP {\bf 0801} (2008) 035
 [arXiv:0711.0786 [gr-qc]]. 
  
\bibitem{Bina:2007wj}
  A.~Bina, K.~Atazadeh and S.~Jalalzadeh,
  Int.\ J.\ Theor.\ Phys.\  {\bf 47} (2008) 1354
  [arXiv:0709.3623 [gr-qc]].

\bibitem{Battisti:2007zg}
  M.~V.~Battisti and G.~Montani,
  Phys.\ Rev.\  D {\bf 77} (2008) 023518
  [arXiv:0707.2726 [gr-qc]].

\bibitem{Zhu:2008cg}
  T.~Zhu, J.~R.~Ren and M.~F.~Li,
  Phys.\ Lett.\  B {\bf 674} (2009) 204
  [arXiv:0811.0212 [hep-th]].

\bibitem{Battisti:2008rv}
  M.~V.~Battisti and G.~Montani,
  Int.\ J.\ Mod.\ Phys.\  A {\bf 23} (2008) 1257
  [arXiv:0802.0688 [gr-qc]].

\bibitem{Vakili:2008zg}
  B.~Vakili,
  Phys.\ Rev.\  D {\bf 77} (2008) 044023
  [arXiv:0801.2438 [gr-qc]].

\bibitem{Myung:2009gv}
  Y.~S.~Myung,
  Phys.\ Lett.\  B {\bf 681} (2009) 81
  [arXiv:0909.2075 [hep-th]].

\bibitem{Myung:2009ur}
  Y.~S.~Myung,
  Phys.\ Lett.\  B {\bf 679} (2009) 491
  [arXiv:0907.5256 [hep-th]].

\bibitem{Ali:2009zq}
  A.~F.~Ali, S.~Das and E.~C.~Vagenas,
  Phys.\ Lett.\  B {\bf 678} (2009) 497
  [arXiv:0906.5396 [hep-th]].

\bibitem{Farmany:2009zz}
  A.~Farmany, M.~Dehghani, M.~R.~Setare and S.~S.~Mortazavi,
  Phys.\ Lett.\  B {\bf 682} (2009) 114.

\bibitem{Li:2009zz}
  Z.~H.~Li,
  Phys.\ Rev.\  D {\bf 80} (2009) 084013.

\bibitem{Bina:2010ir}
  A.~Bina, S.~Jalalzadeh and A.~Moslehi,
  Phys.\ Rev.\  D {\bf 81} (2010) 023528
  [arXiv:1001.0861 [gr-qc]].

\bibitem{Kim:2010wc}
  W.~Kim, Y.~J.~Park and M.~Yoon,
  Mod.\ Phys.\ Lett.\  A {\bf 25} (2010) 1267
  [arXiv:1003.3287 [gr-qc]].

\bibitem{Ali:2011ap}
  A.~F.~Ali,
  Class.\ Quant.\ Grav.\  {\bf 28 } (2011)  065013.
  [arXiv:1101.4181 [hep-th]].

\bibitem{Das:2008kaa}
  S.~Das and E.~C.~Vagenas,
  Phys.\ Rev.\ Lett.\  {\bf 101} (2008) 221301
  [arXiv:0810.5333 [hep-th]].

\bibitem{Vakili:2008tt}
  B.~Vakili,
  Int.\ J.\ Mod.\ Phys.\  {\bf D18 } (2009)  1059-1071.
  [arXiv:0811.3481 [gr-qc]].

\bibitem{Kim:2007bx}
  W.~Kim, J.~J.~Oh,
  JHEP {\bf 0801 } (2008)  034.
  [arXiv:0709.0581 [hep-th]].

\bibitem{Kober:2010sj}
  M.~Kober,
  Phys.\ Rev.\  D {\bf 82} (2010) 085017
  [arXiv:1008.0154 [physics.gen-ph]].

\bibitem{Weinberg:1967tq}
  S.~Weinberg,
  Phys.\ Rev.\ Lett.\  {\bf 19} (1967) 1264.

\bibitem{Weinberg:1996}
  S.~Weinberg,
  {\it The Quantum Theory of Fields, Volume II:~Modern Applications},
  Cambridge University Press, Cambridge 1996.

\bibitem{Yang:1954}
  C.~N.~Yang and R.~L.~Mills,
  Phys.\ Rev.\ {\bf 96}, 191-195 (1954).

\bibitem{Ramond:1990}
  P.~Ramond,
  {\it``Field Theory A modern Primer'', 
  Westview Press 1990.}
  
\bibitem{Calmet:2001na}
  X.~Calmet, B.~Jurco, P.~Schupp, J.~Wess and M.~Wohlgenannt,
  Eur.\ Phys.\ J.\  C {\bf 23} (2002) 363
  [arXiv:hep-ph/0111115].

\bibitem{Calmet:2003jv}
  X.~Calmet and M.~Wohlgenannt,
  Phys.\ Rev.\  D {\bf 68} (2003) 025016
  [arXiv:hep-ph/0305027].

\bibitem{Calmet:2006zy}
  X.~Calmet,
  Eur.\ Phys.\ J.\  C {\bf 50} (2007) 113
  [arXiv:hep-th/0604030].

\bibitem{Chernodub:2007bz}
  M.~N.~Chernodub, A.~J.~Niemi,
  Phys.\ Rev.\  {\bf D77 } (2008)  127902.
  [arXiv:0709.0586 [hep-ph]].

\end{thebibliography}
\end{document}